\documentclass[twocolumn, superscriptaddress,showpacs,amsmath,amssymb,pra,aps,floatfix]{revtex4-1}

\usepackage[latin1]{inputenc}
\usepackage[american]{babel}
\usepackage[T1]{fontenc}
\usepackage{graphicx}
\usepackage{mathrsfs}
\usepackage{hyperref}
\usepackage{bm, epsfig}

\hypersetup{colorlinks=true, urlcolor=blue, citecolor=blue, filecolor=blue, linkcolor=blue}
\newcommand*{\diff}{\mathop{}\!\mathrm{d}}
\newcommand*{\Real}{\mathop{}\!\mathbf{Re}}

\newcommand{\uimm}{\mathrm{i}}
\newcommand{\eu}{\mathrm{e}}
\newcommand{\daga}{^{\dagger}}
\newcommand{\unit}[1]{\ensuremath{\, \mathrm{#1}}}
\newcommand*{\sinc}{\mathop{}\!\mathrm{sinc}}

\begin{document}

\title{Resonance fluorescence in ultrafast and intense x-ray free-electron-laser pulses}

\author{Stefano~M.~Cavaletto}
\email[Corresponding author: ]{smcavaletto@gmail.com}
\affiliation{Max-Planck-Institut f\"{u}r Kernphysik, Saupfercheckweg 1, 69117 Heidelberg, Germany}
\author{Christian~Buth}
\affiliation{Argonne National Laboratory, Argonne, Illinois 60439, USA}
\author{Zolt\'{a}n~Harman}
\affiliation{Max-Planck-Institut f\"{u}r Kernphysik, Saupfercheckweg 1, 69117 Heidelberg, Germany}
\affiliation{ExtreMe Matter Institute (EMMI), Planckstrasse 1, 64291 Darmstadt, Germany}
\author{Elliot~P.~Kanter}
\affiliation{Argonne National Laboratory, Argonne, Illinois 60439, USA}
\author{Stephen~H.~Southworth}
\affiliation{Argonne National Laboratory, Argonne, Illinois 60439, USA}
\author{Linda~Young}
\affiliation{Argonne National Laboratory, Argonne, Illinois 60439, USA}
\author{Christoph~H.~Keitel}
\affiliation{Max-Planck-Institut f\"{u}r Kernphysik, Saupfercheckweg 1, 69117 Heidelberg, Germany}

\begin{abstract}
The spectrum of resonance fluorescence is calculated for a two-level system excited by an intense, ultrashort \mbox{x-ray} pulse made available for instance by free-electron lasers such as the Linac Coherent Light Source. We allow for inner-shell hole decay widths and destruction of the system by further photoionization. This two-level description is employed to model neon cations strongly driven by x~rays tuned to the $1s\,2p^{-1}\rightarrow 1s^{-1}\,2p$ transition at 848 \unit{eV}; the x~rays induce Rabi oscillations which are so fast that they compete with Ne $1s$-hole decay. We predict resonance fluorescence spectra for two different scenarios: first, chaotic pulses based on the self-amplified spontaneous emission principle, like those presently generated at x-ray free-electron-laser facilities and, second, Gaussian pulses which will become available in the foreseeable future with self-seeding techniques. As an example of the exciting opportunities derived from the use of seeding methods, we predict, in spite of above obstacles, the possibility to distinguish at \mbox{x-ray} frequencies a clear signature of Rabi flopping in the spectrum of resonance fluorescence.
\end{abstract}


\maketitle


\section{Introduction}

\begin{figure}[b]
\centering
\includegraphics[width=0.8\linewidth, keepaspectratio]{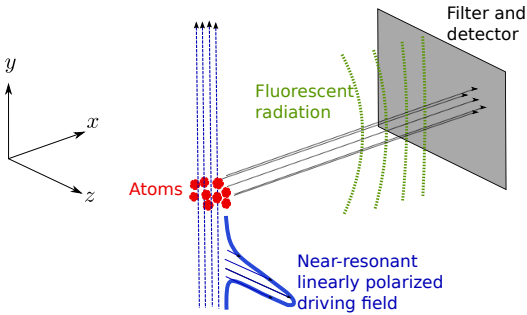}
\caption{(Color online) An atom ensemble (red) is driven by x~rays (blue) tuned to a resonance. The emitted photons (green) are measured perpendicularly to the propagation direction of the x~rays.}
\label{Fig:experimental_setup}
\end{figure}

The spectrum of resonance fluorescence which is emitted by an ensemble of atoms and ions driven by an intense near-resonant electric field \cite{Knight198021, 0034-4885-43-7-002} is one of the cornerstones of quantum optics \cite{Scully:QuantumOptics, Meystre:ElementsOfQuantumOptics}. The spectrum is measured experimentally by exposing an atomic ensemble to intense light and detecting the scattered photons as shown in Fig.~\ref{Fig:experimental_setup}. During the last few decades such studies have received wide attention and have stimulated the development of non-perturbative methods in quantum electrodynamics for the study of the coherent interaction between light and matter \cite{PhysRev.188.1969, PhysRevA.12.1919, PhysRevA.42.1630, PhysRevA.43.3748, PhysRevA.29.2552, PhysRevA.29.2565}. 

The resonance fluorescence spectrum of a two-level system driven by a monochromatic electric field is the simplest case and has been studied extensively at optical frequencies \cite{PhysRev.188.1969, PhysRevA.12.1919, 0022-3700-9-8-007, PhysRevA.13.2123, PhysRevA.15.689}. For a sufficiently strong continuous-wave (cw) driving field a nonlinear three-peak structure appears in the spectrum \cite{0022-3700-7-7-002, PhysRevLett.35.1426, PhysRevA.15.227} which is explained theoretically by the nonperturbative approach of Mollow \cite{PhysRev.188.1969, PhysRevA.12.1919}. The presence of three peaks, frequently called dynamic Stark splitting, is explained as the result of the dressing of bare levels by the external field \cite{0022-3700-10-3-005}.

A cw field is one of the few cases for which an exact analytical solution of the equations of motion (EOMs) of the two-level system exists. When the system interacts with a short pulse, a special class of time-dependent functions, including the case of a hyperbolic secant pulse, were analytically explored for particular values of the physical parameters \cite{PhysRev.40.502, PhysRevA.17.247, PhysRevA.23.2496} and a rich multipeak structure in the spectrum of resonance fluorescence was predicted \cite{0022-3700-17-15-005, PhysRevA.31.1558, Lewenstein:86, PhysRevA.37.1576, PhysRevA.22.2098, PhysRevA.40.3164, PhysRevA.64.013813}. This property, which still represents a signature of Rabi oscillations induced by the intense driving field, is also predicted to depend upon the pulse area, but cannot be intuitively explained by means of dressed states \cite{0022-3700-19-19-004}.

In this paper we investigate, in terms of a two-level model, the coherent interaction of x~rays with core electrons by exciting $K$-shell transitions. Previous studies of strong-field resonance fluorescence have been relevant only at optical frequencies, for which a wide range of models and schemes have been investigated \cite{PhysRevLett.76.388, PhysRevLett.77.3995, PhysRevLett.81.293, PhysRevLett.83.1307, PhysRevLett.91.123601, Ficek_book, Kiffner_review}, because of the lack of coherent and sufficiently intense light sources at short wavelengths. The recent construction of \mbox{x-ray} free-electron lasers (XFELs) \cite{nphoton.2010.176, nphoton.2007.76, nphoton.2011.178, RepXFEL, Galayda:10} provides one with tunable \mbox{x-ray} pulses of unprecedented brilliance, up to one billion times higher than the intensity available at third-generation synchrotron facilities. The intense and ultrafast pulses now available at XFELs offer the opportunity to study nonlinear physics at short wavelengths \cite{VinkoNature, Nature.466.56, PhysRevLett.106.083002, PhysRevLett.104.253002, *Buth:UA-12, PhysRevLett.105.083004, PhysRevLett.105.083005, PhysRevLett.98.183001, 0953-4075-43-19-194008, PhysRevLett.106.033001, PhysRevLett.105.183001}.  In the particular case that we are going to investigate here, x~rays are able to induce stimulated emission and absorption of photons (Rabi flopping) at a time scale that can be compared to and, therefore, compete with the ultrafast inner-shell Auger decay \cite{PhysRevA.77.053404, PhysRevLett.107.233001}.

Existing facilities such as the Linac Coherent Light Source (LCLS) are based on the principle of self-amplified spontaneous emission (SASE) \cite{DoklAkadNauk, *SovPhysDokl, OptComm50.373}, i.e., the beam shot noise gives rise to the emitted radiation which, as a result, possesses only partial temporal coherence and a spiky temporal profile. An analogous situation occurred at the beginning of optical laser science, when it was timely to study the interaction with the chaotic pulses available at that time \cite{PhysRevLett.42.1609, Vannucci:80}. Self-seeding or optical laser-seeding methods are being developed, for which the emitted light is produced by the amplification of a regular (Gaussian) seeding pulse, which exhibits high temporal coherence \cite{Feldhaus1997341, *Saldin2001357, arXiv:1003.2548v1, *arXiv:1008.3036v1, Amann}. The rapid development of XFEL sources makes, therefore, further theoretical work timely \cite{Bern_paper}.

In a recent experiment, intense and ultrashort \mbox{x-ray} pulses from the LCLS have been used to excite the $1s\,2p^{-1}\rightarrow 1s^{-1}\,2p$ transition at 848 \unit{eV} in $\text{Ne}^+$ \cite{PhysRevLett.107.233001}. The electron spectrum of resonant Auger decay was measured to investigate Rabi flopping. With only partial coherence of the SASE pulses presently available at LCLS and the lack of means for single-shot diagnostics, though, the clear observation of Rabi oscillations and its distinction from noise effects is challenging \cite{PhysRevLett.107.233001}. 

Auger decay is the predominant mechanism of inner-shell hole decay of light elements such as neon. Because of the large Auger yield, high-resolution electron spectroscopy is well suited for soft x~rays. Besides Auger decay, \mbox{$K$ holes} also decay by \mbox{x-ray} fluorescence, i.e., by spontaneous emission of photons, while the system is driven by an external field. The spectrum of resonance fluorescence represents an alternative way to study the coherent and nonlinear interaction between x~rays and atoms and ions. It complements the results coming from the detection of electron spectra of resonant Auger decay. High gas densities can generally be used for \mbox{x-ray} emission spectroscopy---orders of magnitude larger than with electron spectroscopy---which compensates for the small fluorescence yield and enables high-resolution measurements with gratings or crystal spectrometers. In the case of resonance fluorescence, however, self-absorption can produce line broadening \cite{corney2006atomic}, so the gas density and path length need to be adjusted to minimize self-absorption effects and make them negligible. Furthermore, photons are scattered much less off electrons or ions in the interaction volume than electrons, i.e., space-charge effects are of little concern \cite{PhysRevLett.105.043003}. 

For \mbox{x-ray} energies, present instruments are expected to detect the fluorescence spectrum with high frequency resolution. For the purposes of this paper, there are at least three choices of instruments: a cryogenic spectrometer \cite{RevModPhys.75.1243}, a diffraction grating \cite{Agren197827} and a crystal spectrometer \cite{PhysRevA.4.476, *1402-4896-7-4-006}. Cryogenic spectrometers such as microcalorimeters are mostly used in the astrophysics community for detecting x~rays from atoms with high atomic numbers and high fluorescence yield \cite{Silver2005683}; at 1 \unit{keV}, Ref.~\cite{refId0} suggests a frequency resolution lower than 0.8 \unit{eV}. For modern grating instruments based on the design described in \cite{nordgren:1690} and \cite{1009-0630-12-3-21} a resolution of 0.4 \unit{eV} at 848 \unit{eV} is expected. It might also be possible to achieve higher frequency resolution by using higher-order reflections from gratings and crystals---though with loss of detection efficiency. With the use of wavelength-dispersive spectrometers, such as diffraction gratings and crystal spectrometers, one might take advantage of their polarization sensitivity in a parallel and perpendicular setup, e.g., for background reduction \cite{brennan:2243}.

In this paper, we develop a time-dependent theory of resonance fluorescence to study the interaction of a two-level model with \mbox{x-ray} pulses. In Sec.~\ref{Theoretical model} we describe our theoretical approach, by defining the spectrum of resonance fluorescence and its main properties and by introducing the two-level model that is used throughout the paper. Results are discussed in Sec.~\ref{Discussion}, where the spectrum of resonance fluorescence is examined for different driving pulses. In particular, we compare different spectra for the presently available chaotic pulses produced via the SASE principle and for pulses with a Gaussian temporal profile that seeding techniques are making available. Section \ref{Conclusion} concludes the paper. Atomic units are used throughout unless otherwise stated.


\section{Theoretical model}
\label{Theoretical model}

\subsection{Two-level model}
\label{Two-level model}

\begin{figure}[tb]
\centering%
\includegraphics[width=0.95\linewidth, keepaspectratio]{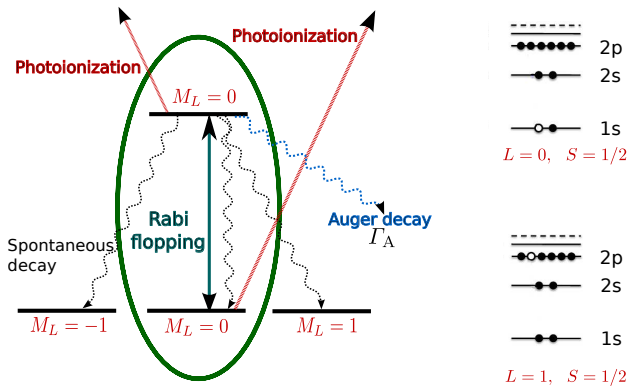}
\caption{(Color online) Two-level model used to describe the coherent interaction between Ne$^+$ and the external driving field tuned to the $1s \,2p^{-1} \rightarrow 1s^{-1}\, 2p$ transition at 848 \unit{eV} \cite{PhysRevLett.107.233001}. The ground state $1s\,2p^{-1}$ is given by $|1\rangle = |L =1,\, M_L = 0\rangle$ and the core-excited state $1s^{-1}\, 2p$ is written as $|2\rangle = |L =0,\, M_L = 0\rangle$. The external field is linearly polarized along the $z$ direction and induces Rabi flopping between these states. Spontaneous decay, however, also allows the core-excited state to decay to valence-ionized states with $M_L = \pm 1$.
}
\label{Fig:themodel}
\end{figure}

The coherent interaction between atoms and ions and x~rays tuned to a particular atomic resonance can be described in terms of a two-level model when the transition is isolated from other levels. In our case, we use such a model to study the $1s\,2p^{-1}_z\rightarrow 1s^{-1}\,2p_z$ transition in Ne$^+$ at an energy of 848 \unit{eV} \cite{PhysRevLett.107.233001}, driven by a near-resonant electric field linearly polarized along the $z$ direction. The two-level model, which is depicted in Fig.~\ref{Fig:themodel}, is justified by the fact that the transition is very well isolated, by more than 70 natural linewidths separated from the next Rydberg excitation $1s\rightarrow 3p$ of neutral Ne at 867 \unit{eV} \cite{PhysRevLett.107.233001}. For neon, relativistic effects and fine-structure splitting do not play an important role and, therefore, spin-orbit splitting can be neglected. 

We describe the emitted fluorescent light field by a quantum operator $\hat{\boldsymbol{E}}(\boldsymbol{r},\, t) = \hat{\boldsymbol{E}}^{+}(\boldsymbol{r},\, t) + \hat{\boldsymbol{E}}^{-}(\boldsymbol{r},\, t)$, where $\hat{\boldsymbol{E}}^{+}(\boldsymbol{r},\, t)$  and $\hat{\boldsymbol{E}}^{-}(\boldsymbol{r},\, t)$ are respectively the positive-frequency and negative-frequency parts of the operator \cite{diels2006ultrashort}. However, it is sufficient to describe the relatively strong driving field classically \cite{PhysRevA.13.2123}, via 
\begin{equation}
\boldsymbol{\mathcal{E}}(t) = \boldsymbol{\mathcal{E}}_0(t) \cos[\omega_{\mathrm{X}}t + \varphi_{\mathrm{X}}(t) + \varphi_{\mathrm{X,0}}], 
\label{eq:classical_field}
\end{equation}
where $\boldsymbol{\mathcal{E}}_0(t)$ is the time-dependent electric-field envelope, $\omega_{\mathrm{X}}$ is the \mbox{x-ray} central frequency, $\varphi_{\mathrm{X}}(t)$ is the time-dependent phase of the field, and $\varphi_{\mathrm{X,0}}$ is the carrier-envelope phase (CEP). We assume throughout an electric field linearly polarized along the $z$ direction, $\boldsymbol{\mathcal{E}}_0(t) = {\mathcal{E}}_0(t)\, \hat{\boldsymbol{e}}_z$, with the unit vector in $z$ direction $\hat{\boldsymbol{e}}_z$. The use of planar undulators at XFEL facilities, in fact, produces linearly polarized \mbox{x-ray} pulses \cite{Galayda:10}; experimental evidence for a very high degree of linear polarization of LCLS x~rays has been given in Refs. \cite{Nature.466.56, PhysRevLett.107.233001}. We further assume a pulse with uniform intensity distribution profile; spatial averaging is therefore not performed.

In order to properly model the atomic transition, we see in Fig.~\ref{Fig:themodel} that the $1s\, 2p^{-1}$ configuration is a spin doublet state with a total orbital angular momentum of $L = 1$; consequently, it is triply degenerate in energy. The three eigenstates of the unperturbed atomic Hamiltonian $\hat{H}_0$ with energy $\omega_{1}$, which diagonalize the $z$ component of the total orbital angular momentum operator, are $|1_{+}\rangle $, $|1_0\rangle $ and $|1_{-}\rangle $, respectively with $M_L = +1,\,0,\,-1$. Conversely, the $1s^{-1}\,2p$ configuration corresponds to the single eigenstate $|2\rangle$ of the field-free atomic Hamiltonian, with $L=0$, $M_L=0$, and energy $\omega_2$. The energy of the atomic transition is $\omega_{21} = \omega_2 - \omega_1$. The relevant raising and lowering atomic operators are
\begin{equation}
\hat{\sigma}_{ij} = |i\rangle\langle j|, \ \ i,\,j \in \{1_+,\,1_0,\,1_-,\,2\}.
\label{eq:raiselower}
\end{equation}

The interaction between the ions and the electric field is described in the dipole approximation because the $1s$ orbital of neon is very compact, involving dimensions much smaller than the wavelength associated with the transition $1s\,2p^{-1}\rightarrow 1s^{-1}\,2p$,  such that nondipole terms are small \cite{PhysRevA.77.053404}. The Hamiltonian of the system is 
\begin{equation}
\hat{H} = \hat{H}_0 + \hat{H}_{\mathrm{int}},
\label{eq:totalHamiltonian}
\end{equation}
where $\hat{H}_0 = \sum_i{\omega_i\,\hat{\sigma}_{ii}}$ is the unperturbed atomic Hamiltonian with eigenvalues $\omega_{i}$, whereas $\hat{H}_{\mathrm{int}}$ represents the interaction of the ion with the classical, linearly polarized, near-resonant field (\ref{eq:classical_field}) \cite{Scully:QuantumOptics},
\begin{equation}
\hat{H}_{\mathrm{int}} = - \hat{\boldsymbol{P}}\cdot \boldsymbol{\mathcal{E}}_0(t)\cos[\omega_{\mathrm{X}} t + \varphi_{\mathrm{X}}(t) + \varphi_{\mathrm{X,0}}].
\label{eq:firts_interaction_hamiltonian}
\end{equation}
The operator $\hat{\boldsymbol{P}}$ in (\ref{eq:firts_interaction_hamiltonian}) is the total atomic polarization operator
\begin{equation}
\hat{\boldsymbol{P}} = \hat{\boldsymbol{P}}^{+} + \hat{\boldsymbol{P}}^{-},
\label{eq:polarization}
\end{equation}
with $\hat{\boldsymbol{P}}^{-} = [\hat{\boldsymbol{P}}^{+}]\daga$ and \cite{Foot:AtomicPhysics}
\begin{equation}
\begin{split}
\hat{\boldsymbol{P}}^{+} & = \langle 1_+|\hat{\boldsymbol{P}}|2\rangle\,\hat{\sigma}_{1_{+}2}+ \langle 1_0|\hat{\boldsymbol{P}}|2\rangle \,\hat{\sigma}_{1_0 2}+ \langle 1_{-}|\hat{\boldsymbol{P}}|2\rangle \,\hat{\sigma}_{1_{-} 2}\\
& =\wp  ( \hat{\boldsymbol{e}}_{\sigma^+} \,\hat{\sigma}_{1_{+}2}+ \hat{\boldsymbol{e}}_z \,\hat{\sigma}_{1_02}+ \hat{\boldsymbol{e}}_{\sigma^-} \,\hat{\sigma}_{1_{-}2}),
\end{split}
\label{eq:polarization_in_model}
\end{equation}
where $\hat{\boldsymbol{e}}_x$, $\hat{\boldsymbol{e}}_y$, and $\hat{\boldsymbol{e}}_z$ are unit vectors in the $x$, $y$, and $z$ directions and $\hat{\boldsymbol{e}}_{\sigma^{\pm}} = (\mp \hat{\boldsymbol{e}}_x + \uimm \hat{\boldsymbol{e}}_y)/\sqrt{2}$ are circular polarization vectors, positive (negative) for polarizations $\lambda = \pm1$, with $\hat{\boldsymbol{e}}_{\sigma^{\pm}} = - \hat{\boldsymbol{e}}_{\sigma^{\mp}}^* $. Due to spherical symmetry, the dipole matrix element is real, $\wp = \wp^*$, and is the same for all transitions; as the atomic states have a definite parity, $\langle 1_{\pm,\,0} |\hat{\boldsymbol{P}}|1_{\pm,\,0}\rangle =\langle2|\hat{\boldsymbol{P}}|2\rangle=0 $, dipole transitions only couple states with different total angular momentum quantum numbers, $\Delta L = 1$ \cite{Foot:AtomicPhysics}.

Since the external electric field is assumed to be linearly polarized \cite{Nature.466.56, PhysRevLett.107.233001}, the dipole interaction only couples the states $|2\rangle$ and $|1_0\rangle$ satisfying the condition $\Delta M_L =0$ \cite{Foot:AtomicPhysics}: in Eq.~(\ref{eq:firts_interaction_hamiltonian}), within the rotating-wave approximation \cite{Scully:QuantumOptics} and by using Eq.~(\ref{eq:polarization_in_model}), $\hat{H}_{\mathrm{int}}$ reduces to
\begin{equation}
\hat{H}_{\mathrm{int}} = -\frac{\varOmega_{\mathrm{R}}(t)}{2}\left(\hat{\sigma}_{1_02}\,\eu^{\uimm [\omega_{\mathrm{X}}t + \varphi_{\mathrm{X}}(t)]} + \hat{\sigma}_{21_0}\,\eu^{-\uimm [\omega_{\mathrm{X}}t + \varphi_{\mathrm{X}}(t)]}\right),
\label{eq:interaction_hamiltonian}
\end{equation}
where we have set the CEP $\varphi_{\mathrm{X,0}}$ to $0$ and where the instantaneous Rabi frequency 
\begin{equation}
\varOmega_{\mathrm{R}}(t)= \wp\,\mathcal{E}_0(t)
\label{eq:omegaR}
\end{equation}
has been introduced.

In our model the dynamics of the two \mbox{x-ray} coupled states $|2\rangle$ and $|1_0\rangle$ is independent from the other two states $|1_{\pm}\rangle$ and one can develop an actual two-level description of the system in which the EOMs exclusively contain the two aforementioned states $|1_0\rangle=|1\rangle$ and $|2\rangle$ and neglect the other two states entirely.

\subsection{Density matrix formulation and equations of motion}
\label{Density matrix formulation and equations of motion}

We investigate in the following the two-level system formed by the states $|1\rangle \equiv |1_0\rangle$ and $|2\rangle$.

We introduce the density matrix
\begin{equation}
{\rho}_{ij}(t) =\langle i|\hat{\rho}(t)|j\rangle = \langle \hat{\sigma}_{ji}(t)\rangle
\label{eq:density_matrix}
\end{equation}
($i,\,j\in\{1,\,2\}$), whose evolution is described by the master equation \cite{Blum:DensityMatrix}
\begin{equation}
\frac{\diff \hat{\rho}}{\diff t} = -\uimm\,[\hat{H}, \hat{\rho}(t)] + \mathcal{L}\hat{\rho}(t) + \mathcal{D}\hat{\rho}(t).
\label{eq:master_eq}
\end{equation}
The first term $-{\uimm}\,[\hat{H}, \hat{\rho}(t)]$ describes the coherent dynamics of the two-level system. In the total Hamiltonian $\hat{H}$ [Eq.~(\ref{eq:totalHamiltonian})] the only relevant terms of the unperturbed atomic Hamiltonian $\hat{H}_0$ are $\omega_1\hat{\sigma}_{11}$ and $\omega_{2}\hat{\sigma}_{22}$. The Lindblad operator $\mathcal{L}\hat{\rho}(t)$ represents the norm-conserving spontaneous decay of the population from the excited state $|2\rangle$ to the ground state $|1\rangle$. The rate at which this process occurs is given by \cite{Scully:QuantumOptics}
\begin{equation}
\varGamma_{\mathrm{R},z} = \frac{4\omega_{21}^3}{3c^3}|\wp|^2.
\label{eq:gammaradz}
\end{equation}
Atoms and ions with high atomic numbers are usually characterized by a high fluorescence yield, i.e., the importance of spontaneous decay increases with the atomic number of the ion of interest.
The last term $\mathcal{D}\hat{\rho}(t)$ denotes the norm-nonconserving term not present in the Lindblad form of the master equation \cite{Louisell:QuantumStatisticalPropertiesOfRadiation}. We introduce this term to describe the decrease of the population of both the upper and lower states \cite{Meystre:ElementsOfQuantumOptics}. These norm-nonconserving processes include Auger decay, photoionization of the system due to the intense external field, and spontaneous decay from the excited level $|2\rangle$ to the two levels $|1_{\pm}\rangle$ which are not coupled by dipole interaction.  We do not include Doppler broadening \cite{corney2006atomic} and collision effects \cite{Meystre:ElementsOfQuantumOptics} in our model, because they involve time scales much longer than the decay time of the system and, at room temperature and for a pressure of 1 atm, they can be neglected \footnote{In the case of neon cations driven on the $1s\,2p^{-1}\rightarrow 1s^{-1}\,2p$ transition at $\omega_{21} = 848\,\unit{eV}$ which we are going to consider in the following, at $300\,\unit{K}$ and for a pressure of 1 atmosphere, the Doppler broadening is $0.0024\,\unit{eV}$, which is much less than the decay width of the system. Using the kinetic theory of gases, the mean time between collisions is approximately $1\,\unit{ns}$, longer by many orders of magnitude than the lifetime of a $1s$ hole.}.

Auger decay and photoionization destroy the two-level system by further ionization of Ne$^+$ to levels which need not be taken into account explicitly. The Auger decay width is $\varGamma_{\mathrm{A}}$, whereas the rate of photoionization $\varGamma_{\mathrm{P},i}(t)$, $i\in\{1,\,2\}$, is \cite{Als-Nielsen:EM-01}
\begin{equation}
\varGamma_{\mathrm{P},i}(t) = \sigma_{\mathrm{X},i} \mathcal{J}_{\mathrm{X}}(t),
\label{eq:photodistruction}
\end{equation}
with the photoionization cross section for the level $i$ $\sigma_{\mathrm{X},i} = \sigma_i(\omega_{\mathrm{X}})$, the \mbox{x-ray} flux 
\begin{equation}
\mathcal{J}_{\mathrm{X}}(t) = {I(t)}/{\omega_{\mathrm{X}}}
\label{eq:flux}
\end{equation}
and the \mbox{x-ray} intensity 
\begin{equation}
I(t) = \frac{\mathcal{E}^2_0(t)}{8\pi\alpha}.
\label{eq:intensity}
\end{equation}
Notice that we evaluate the photoionization cross section and the flux at $\omega_{\mathrm{X}}$ since the cross sections do not vary much within the bandwidth of the field.

The spontaneous decay of the excited level $|2\rangle$ to the states $|1_{\pm}\rangle$ also represents a process which does not conserve the norm of our two-level system. The total radiative decay rate is given by
\begin{equation}
\varGamma_{\mathrm{R}} =\varGamma_{\mathrm{R,}\sigma^+} +\varGamma_{\mathrm{R,}z} + \varGamma_{\mathrm{R,}\sigma^-} = 3 \varGamma_{\mathrm{R,}z},
\label{eq:gammarad}
\end{equation}
where $\varGamma_{\mathrm{R}, z}$ is the spontaneous decay width to the state $|1\rangle$ given in (\ref{eq:gammaradz}) and $\varGamma_{\mathrm{R,}\sigma^{\pm}}$ are defined analogously for the other two decay channels; the second equality exploits Eq.~(\ref{eq:polarization_in_model}). Since the spontaneous decay of the excited level $|2\rangle$ only depends on the population of the state itself, as we are going to show in the following EOMs, the actual dynamics of states $|1_{\pm}\rangle$ can be indeed neglected for our purposes.

The total decay processes are included in Eq.~(\ref{eq:master_eq}). In order to derive the EOMs for the four relevant components of the density matrix, we move to the rotating frame \cite{0022-3700-13-2-011}, by introducing the operators 
\begin{equation}
\hat{\varsigma}_{ii}=\hat{\sigma}_{ii},\ \ \ \hat{\varsigma}_{12}=\eu^{\uimm \omega_{\mathrm{X}}  t} \, \hat{\sigma}_{12},\ \ \  \hat{\varsigma}_{21}= \eu^{-\uimm \omega_{\mathrm{X}}  t}\,  \hat{\sigma}_{21},
\label{eq:varsigma}
\end{equation}
whose expectation values are denoted by 
\begin{equation}
R_{ij}(t)= \langle\hat{\varsigma}_{ji}(t)\rangle,
\label{eq:Rij}
\end{equation}
which, from (\ref{eq:density_matrix}) and (\ref{eq:varsigma}), implies that $R_{ii}(t)=\rho_{ii}(t) $, $R_{12} = \rho_{12}\,\eu^{-\uimm\omega_{\mathrm{X}}t}$ and $R_{21} = \rho_{21}\,\eu^{\uimm\omega_{\mathrm{X}}t}$.

We introduce the vector 
$$\vec{R}(t)= (R_{11}(t),\, R_{12}(t),\,R_{21}(t),\,R_{22}(t)\,)^{\mathrm{T}},$$
whose components are given by the elements of the density matrix $R_{ij}(t)$ in the rotating frame. Before the arrival of the light pulse the two-level system is assumed to be in the ground state, i.e., $\vec{R}_0 = (1,\, 0,\,0,\,0\,)^{\mathrm{T}}$. This assumption is supported by experimental observations of orbital alignment in ions produced by strong-field ionization \cite{PhysRevLett.97.083601}. If the fraction of ions in the $M_L = 0$ ground state is lower than 1, the resonance fluorescence spectrum must be multiplied by this factor.

The master equation (\ref{eq:master_eq}) can be rewritten in matrix form
\begin{equation}
\frac{\diff \vec{R}(t)}{\diff t} = \mathbf{M}(t)\vec{R}(t),\ \ \ \vec{R}(0)=\vec{R}_0,
\label{eq:diffR}
\end{equation}
where $\mathbf{M}(t)$ is the following time-dependent matrix
\begin{widetext}
\begin{equation}
\mathbf{M}(t) = 
\begin{pmatrix}
-\gamma_1(t) &-\uimm \frac{\varOmega_{\mathrm{R}}(t)}{2}\,\eu^{-\uimm \varphi_{\mathrm{X}}(t)} & \uimm \frac{\varOmega_{\mathrm{R}}(t)}{2}\,\eu^{\uimm \varphi_{\mathrm{X}}(t)} & +\varGamma_{\mathrm{R},z}\\
-\uimm\frac{\varOmega_{\mathrm{R}}(t)}{2}\,\eu^{\uimm \varphi_{\mathrm{X}}(t)} &\uimm\varDelta -\frac{1}{2}\left(\gamma_1(t) + \gamma_2(t)\right) &0 &\uimm \frac{\varOmega_{\mathrm{R}}(t)}{2}\,\eu^{\uimm \varphi_{\mathrm{X}}(t)} \\
\uimm\frac{\varOmega_{\mathrm{R}}(t)}{2}\,\eu^{-\uimm \varphi_{\mathrm{X}}(t)} &0 &-\uimm\varDelta -\frac{1}{2}\left(\gamma_1(t) + \gamma_2(t)\right) &-\uimm \frac{\varOmega_{\mathrm{R}}(t)}{2}\,\eu^{-\uimm \varphi_{\mathrm{X}}(t)} \\
0  &\uimm \frac{\varOmega_{\mathrm{R}}(t)}{2}\,\eu^{-\uimm \varphi_{\mathrm{X}}(t)} & -\uimm \frac{\varOmega_{\mathrm{R}}(t)}{2}\,\eu^{\uimm \varphi_{\mathrm{X}}(t)} &-\gamma_2(t)
\end{pmatrix},
\end{equation}
\end{widetext}
with 
\begin{subequations}
\begin{align}
\gamma_1(t) & = \sigma_{\mathrm{X},1} \mathcal{J}_{\mathrm{X}}(t), \\
\gamma_2(t) & = \sigma_{\mathrm{X},2} \mathcal{J}_{\mathrm{X}}(t) +\varGamma_{\mathrm{A}} + \varGamma_{\mathrm{R}},
\end{align}
\end{subequations}
where we have defined the detuning $\varDelta = \omega_{21} - \omega_{\mathrm{X}}$. 

The knowledge of the time evolution of the atomic one-time expectation values is used to derive the two-time expectation values necessary for the computation of the spectrum of resonance fluorescence. For this purpose, we introduce the two-time vector 
\begin{equation}
\begin{split}
   & \vec{Y}(t_1,t_2) \\ 
=\ & (Y_{11}(t_1,t_2),\, Y_{12}(t_1,t_2),\,Y_{21}(t_1,t_2),\,Y_{22}(t_1,t_2)\,)^{\mathrm{T}},
\end{split}
\end{equation}
whose elements are defined as 
\begin{equation}
Y_{ij}(t_1,t_2)= \langle\hat{\varsigma}_{ji}(t_1)\hat{\varsigma}_{12}(t_2)\rangle. 
\end{equation}
Applying the quantum regression theorem \cite{PhysRev.129.2342, 0022-3700-13-2-011} yields
\begin{equation}
\frac{\partial \vec{Y}(t_1,t_2)}{\partial t_1} = \mathbf{M}(t_1) \vec{Y}(t_1, t_2), \ \ \ t_1\geq t_2,
\label{eq:diffY}
\end{equation}
with the initial conditions given by
$Y_{ij}(t_2,t_2)=\delta_{i1}R_{2j}(t_2)$.
The solution of the first set of differential equations (\ref{eq:diffR}) provides one with the initial conditions for the second set of differential equations (\ref{eq:diffY}), whose solution gives
\begin{equation}
\begin{split}
{Y}_{12}(t_1,\,t_2) & = \langle\hat{\varsigma}_{21}(t_1)\hat{\varsigma}_{12}(t_2)\rangle \\
		  & = \langle\hat{\sigma}_{21}(t_1)\hat{\sigma}_{12}(t_2)\rangle\,\eu^{-\uimm\omega_{\mathrm{X}}(t_1 - t_2)}.
\end{split}
\label{eq:particularY}
\end{equation}


\subsection{Spectrum of resonance fluorescence}
\label{Spectrum of resonance fluorescence}

The study and computation of the spectral properties of the fluorescent light requires the knowledge of the first-order autocorrelation function of the electric-field operator \cite{GlauberLesHouches, PhysRev.130.2529}
\begin{equation}
G^{(1)}(t_1,t_2, \boldsymbol{r}) =  \langle \hat{\boldsymbol{E}}^{-}(\boldsymbol{r},\, t_1)\cdot \hat{\boldsymbol{E}}^{+}(\boldsymbol{r},\, t_2)\rangle.
\label{eq:autocorrelation}
\end{equation}

In the case of cw light, when the first-order autocorrelation function depends explicitly on the time difference $\tau = t_1 - t_2$, i.e., $G^{(1)}(t_1,t_2, \boldsymbol{r}) = G^{(1)}(\tau, \boldsymbol{r})$, the Wiener-Khintchine theorem \cite{Scully:QuantumOptics} states that the power spectrum of resonance fluorescence associated with the rate of photons emitted at a given frequency is well defined and given by the Fourier transform of $G^{(1)}(\tau, \boldsymbol{r})$ \cite{JD198347}. However, for ultrashort light pulses, $G^{(1)}(t_1,t_2, \boldsymbol{r})$ explicitly depends upon the two distinct instants $t_1$ and $t_2$ and the Wiener-Khintchine theorem cannot be analogously used to define a power spectrum. Instead, one needs to study the energy spectrum of resonance fluorescence, defined as a quantity proportional to the probability that an ideal photon detector---modeled itself as a two-level system with tunable transition energy $\omega$---is excited by the fluorescent light. In first order of perturbation theory and in the electric-dipole approximation, the energy spectrum is defined as \cite{GlauberLesHouches}
\begin{equation}
S(\omega,\,\boldsymbol{r}) = \frac{1}{4\pi\alpha}\,\int_{-\infty}^{+\infty}{\int_{-\infty}^{+\infty}{{G^{(1)}(t_1,t_2, \boldsymbol{r})}\,\eu^{-\uimm\omega(t_1-t_2)} \diff t_1} \diff t_2}.
\label{eq:energy_spectrum_definition}
\end{equation}
Here, $S(\omega,\,\boldsymbol{r})\diff\omega\diff A$ represents the energy detected on average in the differential energy interval $[\omega,\,\omega + \diff \omega]$ and in a surface element $\diff \boldsymbol{A} = r^2\,\diff\varOmega\,\hat{\boldsymbol{e}}_r$ centered on $\boldsymbol{r} = r\,\hat{\boldsymbol{e}}_r$. Further, $\alpha$ is the fine-structure constant, $\diff\varOmega$ is the differential solid angle, and $\hat{\boldsymbol{e}}_r = \boldsymbol{r}/|\boldsymbol{r}|$ is the unit vector in the direction of observation ($\boldsymbol{0}$ is the position of the atom).

We assume that the driving field propagates along the \mbox{$y$ axis}. In the far-field limit and in the electric-dipole approximation---away from the $y$ propagation axis in which also the driving field would be present---the electric-field operator associated with the fluorescent light can be related to the atomic polarization operator $\hat{\boldsymbol{P}}^{+}(t)$ [Eq.~(\ref{eq:polarization_in_model})] via the relation \cite{PhysRevA.13.2123, PhysRevA.42.1630, Scully:QuantumOptics}
\begin{equation}
\hat{\boldsymbol{E}}^{+}(\boldsymbol{r}, \,t)= \frac{ {\omega}_{21}^2}{c^2 r}\,\left\{\hat{\boldsymbol{P}}^{+}(t-r/c)- \left[\hat{\boldsymbol{P}}^{+}(t-r/c) \cdot \hat{\boldsymbol{e}}_r\right]\,\hat{\boldsymbol{e}}_r\right\}.
\label{eq:emitted_field}
\end{equation}
If the detector is placed along the \mbox{$x$ axis}, as shown in Fig.~\ref{Fig:experimental_setup}, then $\hat{\boldsymbol{e}}_r = \hat{\boldsymbol{e}}_x$ and one obtains from (\ref{eq:polarization_in_model})
\begin{equation}
\hat{\boldsymbol{E}}^{+}(r\,\hat{\boldsymbol{e}}_x, \,t)= \hat{{E}}^{+}_z(r\,\hat{\boldsymbol{e}}_x, \,t)\,\hat{\boldsymbol{e}}_z + \hat{{E}}^{+}_y(r\,\hat{\boldsymbol{e}}_x, \,t)\,\hat{\boldsymbol{e}}_y,
\end{equation}
with 
\begin{equation}
\hat{{E}}^{+}_z(r\,\hat{\boldsymbol{e}}_x, \,t) = \frac{\wp\,\omega_{21}^2}{c^2 r }\,\hat{\sigma}_{1_02}(t-r/c)
\label{eq:Ez+}
\end{equation}
and
\begin{equation}
\hat{{E}}^{+}_y(r\,\hat{\boldsymbol{e}}_x, \,t) = \frac{\uimm}{\sqrt{2}}\,\frac{\wp\,\omega_{21}^2}{c^2 r }\, \left[\hat{\sigma}_{1_+2}(t-r/c)+\hat{\sigma}_{1_-2}(t-r/c)\right],
\label{eq:Ey+}
\end{equation}
whereas, because of the placement of the detector, the $x$ component of the electric-field operator $\hat{{E}}^{+}_x(r\,\hat{\boldsymbol{e}}_x, \,t)$ vanishes. Analogously, the autocorrelation function is split into two parts:
\begin{equation}
\begin{aligned}
G^{(1)}(t_1,t_2, r\,\hat{\boldsymbol{e}}_x) = \,&\langle \hat{{E}}^{-}_z( r\,\hat{\boldsymbol{e}}_x,\, t_1)\,\hat{{E}}^{+}_z( r\,\hat{\boldsymbol{e}}_x,\, t_2)\rangle\, \\
&+\,\langle \hat{{E}}^{-}_y( r\,\hat{\boldsymbol{e}}_x,\, t_1)\,\hat{{E}}^{+}_y( r\,\hat{\boldsymbol{e}}_x,\, t_2)\rangle.
\end{aligned}
\label{eq:autocorr_twoterms}
\end{equation}
The first term in (\ref{eq:autocorr_twoterms}) is the autocorrelation function of the fluorescence photons which are polarized along the $z$ direction; the transition with which they are associated ($|1\rangle \rightarrow |2\rangle$) is driven by the external field, which modulates the polarization operator along the $z$ direction and induces Rabi flopping. The general case of a detector placed in the $x-z$ plane, forming an arbitrary angle $\theta$ with respect to the \mbox{$x$ axis}, is discussed in Appendix~\ref{Polarization effects and measurement geometry}.

With (\ref{eq:energy_spectrum_definition}) and (\ref{eq:autocorr_twoterms}), the resonance fluorescence energy spectrum is also split into two terms $S(\omega,\,r\,\hat{\boldsymbol{e}}_x) = S_z(\omega,\,r\,\hat{\boldsymbol{e}}_x) + S_y(\omega,\,r\,\hat{\boldsymbol{e}}_x)$. The calculation of $S_y(\omega,\,r\,\hat{\boldsymbol{e}}_x)$ goes beyond the two-level approximation we adopt in this paper and requires a complete four-level description of the system. In this paper we describe the appearance of Rabi flopping in the resonance fluorescence spectrum for those photons which are emitted in the transition to the ground state $|1\rangle$. For $\hat{\boldsymbol{e}}_r = \hat{\boldsymbol{e}}_x$ this represents the only contribution in $S_z(\omega,\,r\,\hat{\boldsymbol{e}}_x)$ and its observation can be experimentally realized using a polarization-dependent detection to selectively detect the radiation which is linearly polarized in the $z$
direction, in order to select those fluorescence photons associated with the transition to the state with $M_L = 0$.

Polarization-dependent measurements can be very informative, e.g., they have played an important role for molecules where the valence orbitals can be resolved \cite{PhysRevLett.60.1010, PhysRevLett.67.1098, *Southworth1991304}. Reflections from mirrors, gratings, or crystals at angles that achieve high polarization selectivity at the frequency of the atomic transition involved allow one to measure the polarization of the radiation. The use of wavelength-dispersive spectrometers, which involve reflecting x~rays from a grating or crystal, can provide one with polarization selectivity. Energy-dispersive spectrometers, such as a cryogenic spectrometer, can be polarization sensitive if they are pixelated and the x~rays are hard enough to Compton scatter in the absorber. In addition, polarization-dependent detection in a parallel and perpendicular setup facilitates background reduction.

By exploiting the polarization properties of the emitted light, we can focus on the first component of the first-order autocorrelation function (\ref{eq:autocorr_twoterms}), which is expressed as
\begin{equation}
G^{(1)}_z(t_1,t_2, r\,\hat{\boldsymbol{e}}_x) = \mathcal{I}(r)\,\langle \hat{\sigma}_{21}(t_1-r/c)\,\hat{\sigma}_{12}(t_2-r/c)\rangle,
\label{eq:autocorrelationconIr}
\end{equation}
where 
\begin{equation}
\mathcal{I}(r)=\Bigl(\frac{\omega_{21}^2\,{|\wp|}}{c^2 r }\Bigr)^2
\label{eq:Ir}
\end{equation}
is a factor dependent on the position of observation at which the detector is placed and having the dimension of an intensity \cite{PhysRevA.13.2123}.

By introducing the time delay $\tau = t_1 - t_2$ and noticing that
$
\langle\hat{\sigma}_{21}(t_1)\hat{\sigma}_{12}(t_2)\rangle = \langle\hat{\sigma}_{21}(t_2)\hat{\sigma}_{12}(t_1)\rangle^*
$,
we conclude that knowledge of $\langle\hat{\sigma}_{21}(t_1) \hat{\sigma}_{12}(t_2)\rangle $ in the region $t_1\geq t_2$ (and hence $\tau\geq0$) is sufficient for the calculation of the energy spectrum of resonance fluorescence \cite{0022-3700-13-2-011}. We rewrite (\ref{eq:energy_spectrum_definition}) in compact form as
\begin{equation}
\begin{split}
S_z(\omega,\,r\,\hat{\boldsymbol{e}}_x) = & \frac{3\varGamma_{\mathrm{R},z}\,\omega_{21}}{8\pi\,r^2}\int_{-\infty}^{+\infty}\int_{0}^{+\infty} \Real\left[\eu^{-\uimm\omega\tau}\langle\hat{\sigma}_{21}(t_2+\tau)\right. \\ 
& \times \left.\hat{\sigma}_{12}(t_2)\rangle\right]\diff \tau\, \diff t_2,
\end{split}
\label{eq:energy_spectrum_compact}
\end{equation}
where we use Eqs. (\ref{eq:gammaradz}), (\ref{eq:autocorrelationconIr}) and (\ref{eq:Ir}). As a result, one can use $Y_{12}(t_1,\,t_2)$ from the solution of (\ref{eq:diffY}) to compute the energy spectrum of resonance fluorescence. 

In the following, we are going to compute $S_z(\omega,\,\varOmega) = r^2\,S_z(\omega,\,r\,\hat{\boldsymbol{e}}_x)$ for a detector along the \mbox{$x$ axis}. $S_z(\omega,\,\varOmega)\,\diff\varOmega\diff\omega$ is the energy emitted into $\diff\varOmega$ and $\diff\omega$; in atomic units $S_z(\omega,\,\varOmega)$ has the dimension of 1/sr. Finally, we notice that the total detected energy emitted into $\diff\varOmega$ is
\begin{equation}
\mathscr{E} =\int_{-\infty}^{+\infty}{S_z(\omega,\,\varOmega)\,\diff\omega} = \frac{3\varGamma_{\mathrm{R},z}\,\omega_{21}}{8}\int_{-\infty}^{+\infty}{R_{22}(t)\,\diff t},
\label{eq:totalEmittedEnergy}
\end{equation}
exploiting the relation $$2\pi\,\delta(t_1- t_2) = \int_{-\infty}^{+\infty}\eu^{-\uimm\omega(t_1-t_2)}\,\diff\omega.$$


\section{Results and discussion}
\label{Discussion}

Here we apply our two-level model to study neon cations on the $1s\,2p^{-1}\rightarrow 1s^{-1}\,2p$ transition at $\omega_{21} = 848\,\unit{eV}$ \cite{PhysRevLett.107.233001}, i.e., the detuning is $\varDelta = \omega_{21} - \omega_{\mathrm{X}} = 0$. Scattered \mbox{x rays} could be observed if the XFEL beam energy is detuned from resonance. As demonstrated in Ref. \cite{PhysRevA.27.923}, for example, Compton scattering, resonant Raman scattering, and Rayleigh scattering can be observed as the resonance is approached from below. At $848\,\unit{eV}$, however, resonance fluorescence will dominate the measured spectrum.

The destruction rate of our effective two-level system is dominated by the Auger decay width of Ne $1s^{-1}$ which is $\varGamma_{\mathrm{A}} = 0.27\,\unit{eV}$ \cite{Schmidt:values}. The dipole moment $\wp = 0.0524\,a_0$ is computed with the Hartree-Fock-Slater mean-field model \cite{HartreeMethod, *springerlink:10.1007/BF01340294,  *PhysRev.81.385, HermanSkillman, FELLA}, whereas the photoionization cross sections are computed using the Los Alamos Atomic Physics Codes \cite{LANL, Cowan:TheoryAtomicSpectra}. From Eq.~(\ref{eq:gammaradz}) the radiative decay width follows, where $\varGamma_{\mathrm{R}, z} = 0.0012\,\unit{eV}$ and the total decay rate is $\varGamma_{\mathrm{R}} = 0.0039\,\unit{eV}$ \cite{PhysRevA.8.649}, in good agreement with Eq.~(\ref{eq:gammarad}). 

The spectrum $S_z(\omega,\,\varOmega)$ that we will compute represents the emitted photons linearly polarized along the $z$ direction from our two-level model. Off-resonant Rayleigh scattering from $2s$ and $2p$ electrons in Ne is, however, not taken into account. This elastic scattering is predicted to be anisotropically distributed for a linearly polarized electric field: in our case, $\boldsymbol{\mathcal{E}}(t) = {\mathcal{E}}(t)\,\hat{\boldsymbol{e}}_z$, the intensity of the elastic scattering would be affected by the source-dependent polarization factor $\sin^2\psi$ \cite{Als-Nielsen:EM-01}, where $\psi$ is the angle between the \mbox{$z$ axis} and the direction of detection $\hat{\boldsymbol{e}}_r$ at which the detector is placed. This additional contribution is not included in the two-level approximation that we implement in this paper. Its only effect is an enlargement of the central peak of the spectrum.

\subsection{Gaussian x-ray pulses}
\label{Gaussian x rays}

Self-seeding techniques at LCLS are providing one with pulses with an approximately Gaussian temporal profile \cite{arXiv:1003.2548v1, *arXiv:1008.3036v1, Amann}; it is interesting therefore to predict the evolution of the atomic properties in time and the spectrum of resonance fluorescence for $\text{Ne}^+$ cations for this case. We write the Gaussian pulse as
\begin{equation}
\mathcal{E}_{0\mathrm{,G}}(t) = \mathcal{E}_{\mathrm{max}}\,\eu^{-[t^2/(2T^2)]},\ \ \varphi_{\mathrm{X}}(t)=0,
\label{eq:Gaussian_function}
\end{equation} 
where $T = \tau_{\mathrm{G}} /( 2\sqrt{\ln{2}})$ and $\tau_{\mathrm{G}}$ is the FWHM of $\mathcal{E}_{0\mathrm{,G}}^2(t)$. The FWHM of $|\tilde{\mathcal{E}}_{0\mathrm{,G}}(\omega)|^2$ is $\Delta\omega_{\mathrm{G}}={4\ln2}/{\tau_{\mathrm{G}}}$, where $$\tilde{\mathcal{E}}_{0\mathrm{,G}}(\omega) =\int{\mathcal{E}_{0\mathrm{,G}}(t)}\,\eu^{\uimm\omega t}\,\diff t = T\,\sqrt{2\pi}\,\mathcal{E}_{\mathrm{max}}\,\eu^{-(\omega^2 T^2/2)}$$ is the Fourier transform of $\mathcal{E}_{0\mathrm{,G}}(t)$ \cite{diels2006ultrashort}. The peak intensity [Eq.~(\ref{eq:intensity})] is $I_{\mathrm{G}} = \mathcal{E}_{\mathrm{max}}^2/(8\pi\alpha)$, yielding a maximum Rabi frequency [Eq.~(\ref{eq:omegaR})] $\varOmega_{\mathrm{RG,max}} = \wp \mathcal{E}_{\mathrm{max}} = \wp \sqrt{8\pi\alpha\,I_{\mathrm{G}}}$. 

Further, we introduce the pulse area
\begin{equation}
Q = \int_{-\infty}^{+\infty}\varOmega_{\mathrm{R}}(t)\,\diff t,
\label{eq:area}
\end{equation}
which was shown to play an important role in the description of the dynamics of a two-level system in interaction with a regular pulse [$\varphi_{\mathrm{X}}(t)=0$] and in the properties of the corresponding resonance fluorescence spectrum \cite{0022-3700-17-15-005, PhysRevA.31.1558, Lewenstein:86, PhysRevA.37.1576}. Let us assume for now that level decay and photoionization are both negligible. Then, for $\varDelta = 0$, if $n$ is a natural number and $Q = 2\pi n$, the final population after the interaction with the pulse is in the ground state, whereas for $Q = 2\pi(n+1/2)$ a complete inversion happens and the total final population occupies the excited state.
For a Gaussian regular pulse the area (\ref{eq:area}) is $Q_{\mathrm{G}} = \varOmega_{\mathrm{RG,max}}\,\tau_{\mathrm{G}}\sqrt{\pi/(2\ln{2})}$.

\begin{figure}[t]
\centering%
\includegraphics[width=0.95\linewidth, keepaspectratio]{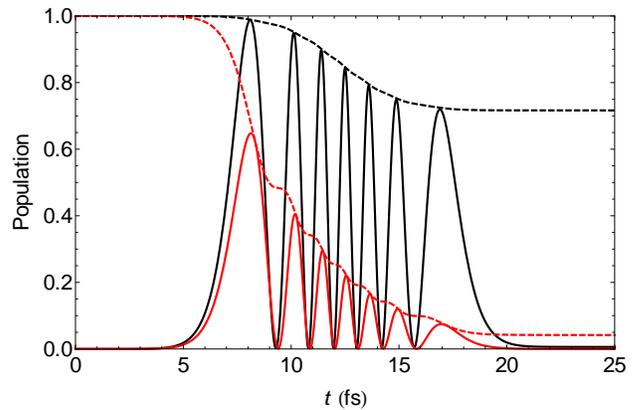}
\caption{(Color online) Time evolution of the population of a two-level system driven by a Gaussian \mbox{x-ray} pulse (\ref{eq:Gaussian_function}) of peak intensity $I_{\mathrm{G}} = 2.6\times 10^{17}\,\mathrm{W/cm^2}$ and FWHM duration $\tau_{\mathrm{G}} = 5\,\unit{fs}$, corresponding to a pulse area of $Q_{\mathrm{G}} = 14\pi$. The dashed lines show the total population of the two-level system $\rho_{11}(t) + \rho_{22}(t)$ [Eq.~(\ref{eq:density_matrix})] in the absence (black line) and presence (red line) of Auger decay. The solid lines show the corresponding occupation of the excited state $\rho_{22}(t)$.}
\label{Fig:regular_pulse_time}
\end{figure}

We begin by studying the interaction of $\text{Ne}^+$ cations driven by a Gaussian \mbox{x-ray} pulse with peak intensity $I_{\mathrm{G}} = 2.6\times 10^{17}\,\mathrm{W/cm^2}$ and $\tau_{\mathrm{G}} = 5\,\unit{fs}$: such \mbox{x-ray} pulses will be available in the foreseeable future from seeding techniques implemented at LCLS. In Fig.~\ref{Fig:regular_pulse_time} we show the time evolution of the two-level system when Auger decay is included and when it is not included ($\varGamma_{\mathrm{A}} = 0$) in the EOMs (\ref{eq:diffR}). The time evolution of the total population of the system reveals that Auger decay is the major depopulation mechanism. Photoionization makes, however, also a noticeable contribution at the chosen \mbox{x-ray} intensity; the maximum rates of photoionization are [Eq.~(\ref{eq:photodistruction})] $\varGamma_{\mathrm{P1,max}} = 0.03\,\unit{eV}$ and $\varGamma_{\mathrm{P2, max}} = 0.04\,\unit{eV}$, which is small compared with the Auger decay width $\varGamma_{\mathrm{A}} = 0.27\,\unit{eV}$. In Ref.~\cite{PhysRevA.77.053404} this channel was, therefore, neglected entirely. The decay time associated with Auger decay is approximately given by $\Delta\tau = 1/\varGamma_{\mathrm{A}} = 2.4\,\unit{fs}$. As one notices in Fig.~\ref{Fig:regular_pulse_time}, the total population of the system almost completely vanishes after the pulse of $5\,\unit{fs}$. Whether Auger decay is included or not does not interfere with Rabi oscillations which are clearly discernible; the pulse area $Q_{\mathrm{G}} = 7\times 2 \pi$ results in seven oscillations.

\begin{figure}[t]
\centering%
\includegraphics[width=0.95\linewidth, keepaspectratio]{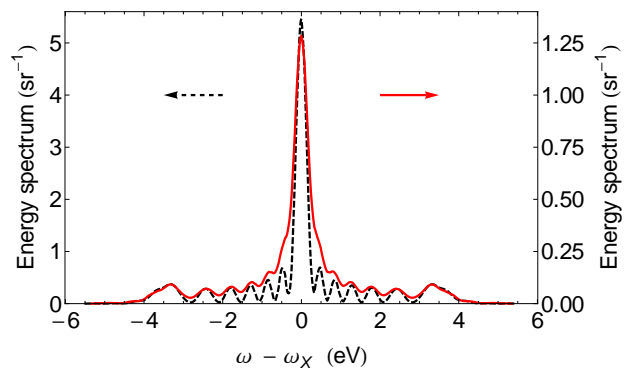}
\caption{(Color online) Energy spectrum of resonance fluorescence for the Gaussian pulse used in Fig.~\ref{Fig:regular_pulse_time} in the absence (black, dashed line) and presence (red, solid line) of Auger decay. As indicated by the arrows, the scale on the left refers to the black, dashed curve, whereas the scale on the right refers to the red, solid curve.}
\label{Fig:regular_spectrum}
\end{figure}

The corresponding energy spectra of resonance fluorescence are shown in Fig.~\ref{Fig:regular_spectrum}. The Rabi oscillations induced by the intense external \mbox{x-ray} field appear in both cases with and without Auger decay, with nonvanishing contributions in the region approximately given by $[-\varOmega_{\mathrm{RG, max}}\,,\,\varOmega_{\mathrm{RG,max}}]$, with the maximum Rabi frequency $\varOmega_{\mathrm{RG, max}} = 3.9\,\unit{eV}$. First, when only spontaneous decay and photoionization are taken into account, a multipeak structure is predicted, in analogy to what was computed in the absence of any decay process \cite{0022-3700-17-15-005}. The presence of many peaks is nontrivially related to the pulse shape of the electric field, i.e., to its finite duration and width. The seven peaks in the energy spectrum---six lateral peaks and the seventh central one---are related, as was shown in \cite{0022-3700-17-15-005}, to the pulse area $Q_{\mathrm{G}} = 7\times 2\pi$. Second, when Auger decay is taken into account, the multipeak structure of the spectrum becomes smoother because of the increase in the decay rate. Furthermore, the intensity of the radiation emitted by the two-level system decreases, since Auger decay destroys it and, consequently, reduces the fraction of atoms which can Rabi flop. The resulting maximum Rabi frequency $\varOmega_{\mathrm{RG, max}} = 3.9\,\unit{eV}$ is however much higher than the bandwidth of the pulse, $\Delta\omega_{\mathrm{G}} = 0.36\,\unit{eV}$, the Auger decay width, $\varGamma_{\mathrm{A}} = 0.27\,\unit{eV}$, and the frequency resolution of present spectrometers, $\Delta\omega_{\mathrm{res}} = 0.4\,\unit{eV}$ \cite{nordgren:1690}. Hence the signature of Rabi flopping, clearly visible in Fig.~\ref{Fig:regular_spectrum}, will be detectable.

\begin{figure}[!tbp]
\centering%
\includegraphics[width=0.915\linewidth, keepaspectratio]{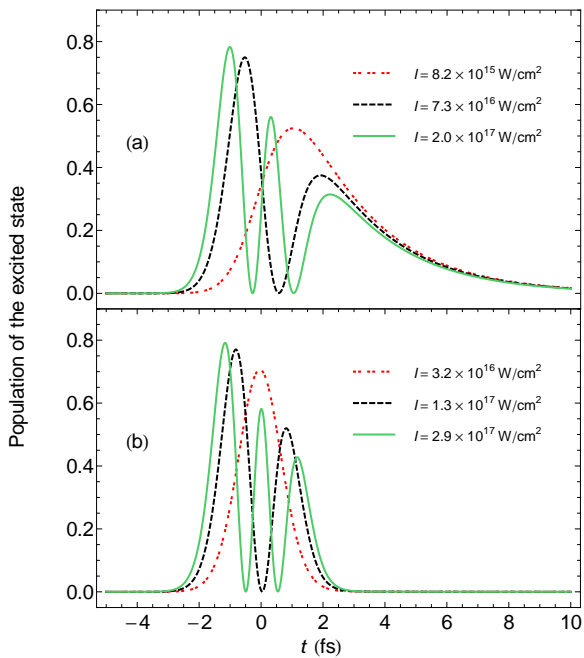}
\caption{(Color online) Time evolution of the population of the excited state $\rho_{22}(t)$ for a two-level system driven by Gaussian \mbox{x-ray} pulses [Eq.~(\ref{eq:Gaussian_function})] of different peak intensities (shown in the legend) and a FWHM duration $\tau_{\mathrm{G}} = 2\,\unit{fs}$. In panel (\textit{a}) pulse areas $Q_{\mathrm{G}} = 2\pi(n-1/2)$, for $n=1$ (red, dotted line), $n=2$ (black, dashed line) and $n=3$ (green, solid line) are used. In panel (\textit{b}) pulse areas $Q_{\mathrm{G}}  = 2\pi n$, for $n=1$ (red, dotted line), $n=2$ (black, dashed line) and $n=3$ (green, solid line) are used.}
\label{Fig:2fsPulserho}
\end{figure}
\begin{figure}[!bp]
\centering%
\includegraphics[width=0.915\linewidth, keepaspectratio]{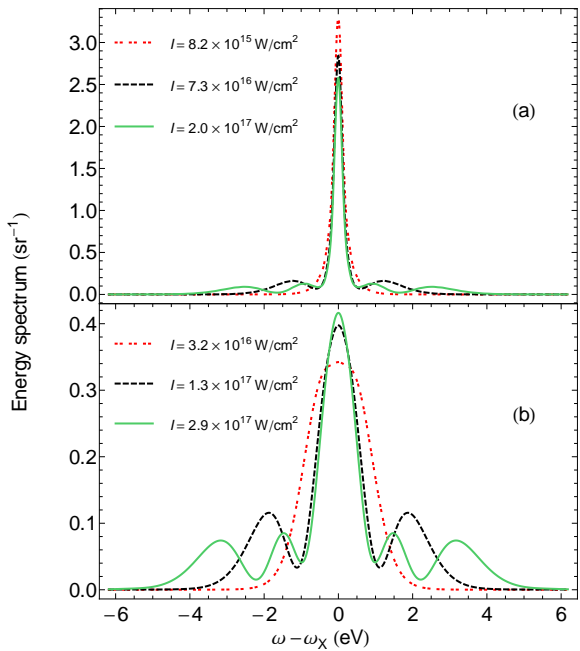}
\caption{(Color online)  Spectrum of resonance fluorescence of a two-level system driven by Gaussian \mbox{x-ray} pulses of different peak intensities (shown in the legend) and a FWHM duration $\tau_{\mathrm{G}} = 2\,\unit{fs}$. Line styles of panels (\textit{a}) and (\textit{b}) as in Fig.~\ref{Fig:2fsPulserho}.}
\label{Fig:2fsPulsesp}
\end{figure}

In Fig.~\ref{Fig:2fsPulserho} and \ref{Fig:2fsPulsesp} we consider different pulses with $\tau_{\mathrm{G}} = 2\,\unit{fs}$ and $Q_{\mathrm{G}} = 2\pi (n-1/2)$ [panels (a)] or $Q_{\mathrm{G}}  = 2\pi n$ [panels (b)], for $n \in \{1,\,2,\,3\}$. One can clearly see a dependence of the population of the two-level system upon the area $Q_{\mathrm{G}}$. When this area is an odd multiple of $\pi$ [Fig.~\ref{Fig:2fsPulserho}\textit{a}], a major part of the population at the end of the pulse occupies the excited state: one can discern the $n-1/2$ oscillations due to the interaction with the pulse and the following Auger decay of the system when the pulse is over. As shown in Fig.~\ref{Fig:2fsPulsesp}\textit{a}, the long Auger decay which follows the interaction with the pulse results in a high Lorentzian peak in the spectrum of resonance fluorescence at $\omega = \omega_{\mathrm{X}}$ with a width that can be related to the major decay process, i.e., $\varGamma_{\mathrm{A}}$. This peak results from the fact that the excited system decays freely, radiatively and electronically, without Rabi flopping. In contrast, a considerably different situation appears when the area of the pulse is an even multiple of $\pi$ [Fig.~\ref{Fig:2fsPulserho}\textit{b}]: after $n$ complete oscillations, the population of the excited state is almost 0 at the end of the pulse. Consequently, the previously present post-x-ray-exposure decay does not take place. The total emitted energy is therefore lower because the central peak at $\omega_{21}$ is reduced by almost one order of magnitude, as one can clearly see by looking at Fig.~\ref{Fig:2fsPulsesp}\textit{b}. In the case of a longer pulse, so that the two-level system has completely Auger decayed before its conclusion, the difference between pulses whose area is an odd or even multiple of $\pi$ becomes less important. 
%

\begin{figure}[!tbp]
\centering%
\includegraphics[width=\linewidth, keepaspectratio]{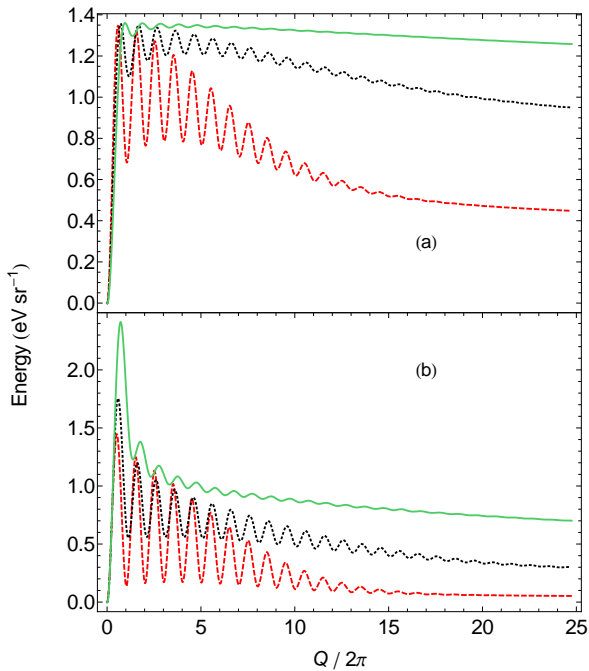}
\caption{(Color online)  (\textit{a}) Total emitted energy $\mathscr{E} = \int_{-\infty}^{+\infty}S_z(\omega,\,\varOmega)\,\diff\omega$ and (\textit{b}) peak value of the spectrum $S(\omega_{\mathrm{X}}) $ multiplied by ${\pi}\varGamma_{\mathrm{A}}/2 $ as functions of the normalized pulse area $Q_{\mathrm{G}}/(2\pi)$ [Eq.~(\ref{eq:area})] for $\tau_{\mathrm{G}} = 2\,\unit{fs}$ (red, dashed line), $\tau_{\mathrm{G}} = 5\,\unit{fs}$ (black, dotted line), and $\tau_{\mathrm{G}} = 10\,\unit{fs}$ (green, solid line).}
\label{Fig:oscillationsArea}
\end{figure}

The dependence of the resonance fluorescence spectrum upon the duration of the pulse is an additional point that needs to be investigated. In order to observe this dependence, in Fig.~\ref{Fig:oscillationsArea} we study the main features of the spectrum as functions of the normalized pulse area $Q_{\mathrm{G}}/(2\pi)$ and of the pulse FWHM duration $\tau_{\mathrm{G}}$. We recall that for fixed $\tau_{\mathrm{G}}$ the area of the Gaussian pulse is directly proportional to the square root of the intensity, $Q_{\mathrm{G}} = 2\pi\wp\tau_{\mathrm{G}} \sqrt{(\alpha/\ln{2})\,I_{\mathrm{G}}}$. 
In Fig.~\ref{Fig:oscillationsArea}\textit{a} we show the total emitted energy $\mathscr{E}$ [Eq.~(\ref{eq:totalEmittedEnergy})] for three different values of $\tau_{\mathrm{G}}$; for the shortest pulses one can clearly observe an oscillating behavior of the total emitted energy as a function of $Q_{\mathrm{G}}/(2\pi)$; this behavior is less pronounced for the longest pulses. It is also worthwhile to notice that, for increasing values of $Q_{\mathrm{G}}$, the intensity can become so high that also for the shortest pulses the system is in any case completely destroyed by photoionization within the duration of the pulse itself. The increasing importance of photoionization implies a less remarkable difference in the time evolution of systems driven by pulses whose area is an odd or even multiple of $\pi$ and, consequently, resonance fluorescence spectra characterized by a lower dependence upon the area of the pulse.

In Fig.~\ref{Fig:oscillationsArea}\textit{b} we display ${\pi\varGamma_{\mathrm{A}}S(\omega_{\mathrm{X}})}/{2} $, where $S(\omega_{\mathrm{X}})$ is the central maximum value of the spectrum of resonance fluorescence. The constant prefactor ${\pi\varGamma_{\mathrm{A}}}/{2}$ allows us to compare the shape of the spectrum of resonance fluorescence with that of a Lorentzian function of Auger decay width $\varGamma_{\mathrm{A}}$. If the only process involved was a decay causing a rate width $\varGamma$, then the spectrum of resonance fluorescence would be a Lorentzian function
\begin{equation}
L(\omega) =  \frac{\pi\varGamma}{2} L_0  \frac{\varGamma/(2\pi)}{(\omega - \omega_{\mathrm{X}})^2 + (\varGamma/2)^2},
\end{equation}
with peak value $L_0 = L(\omega_{\mathrm{X}})$ and with total emitted energy $\mathscr{E}_{L} = \int_{-\infty}^{+\infty}{L(\omega)}\,\diff\omega ={\pi\varGamma L_0}/{2}  $. By computing in Fig.~\ref{Fig:oscillationsArea}\textit{b} the quantity ${\pi\varGamma_{\mathrm{A}}S(\omega_{\mathrm{X}})}/{2} $, we can relate it to the actual total emitted radiation of Fig.~\ref{Fig:oscillationsArea}\textit{a} and understand the relative importance of Auger decay in relation to the other decay processes. By comparing the oscillating features in Fig.~\ref{Fig:oscillationsArea}\textit{b} with those of Fig.~\ref{Fig:oscillationsArea}\textit{a}, one notices that ${\pi\varGamma_{\mathrm{A}}S(\omega_{\mathrm{X}})}/{2} $ approaches $\mathscr{E}$ only for short pulses satisfying $Q_{\mathrm{G}} = 2\pi(n-1/2)$. In these cases, as we have already discussed in Fig.~\ref{Fig:2fsPulserho}\textit{a}, the main term is represented by post-x-ray-exposure Auger decay of the system. Nonetheless, because of the non-negligible role played by Rabi flopping, photoionization and spontaneous decay, one can notice in Fig.~\ref{Fig:oscillationsArea} a clear difference between ${\pi\varGamma_{\mathrm{A}}S(\omega_{\mathrm{X}})}/{2}$ and $\mathscr{E}$.
%

\begin{figure}[bt]
\centering%
\includegraphics[width=0.95\linewidth, keepaspectratio]{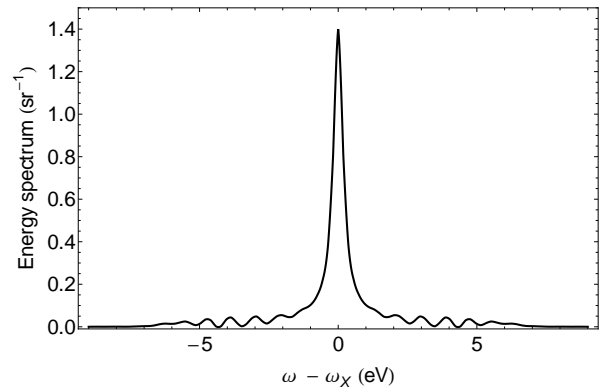}
\caption{Average resonance fluorescence spectrum for 500 different realizations of the Gaussian driving pulses [Eq.~(\ref{eq:Gaussian_function})]. The mean duration of the pulses is $\bar{\tau}_{\mathrm{G}} = 7\,\unit{fs}$, and the mean peak intensity is $\bar{I}_{\mathrm{G}} = 7\times10^{17}\,\mathrm{W/cm^2}$. Here $\tau_{\mathrm{G}}$ and $I_{\mathrm{G}}$ are Gaussian random variables independently chosen for each realization with a variance equal to the $20\%$ of the respective mean values.}
\label{Fig:regular_averaged_spectrum}
\end{figure}

Figures \ref{Fig:regular_spectrum} and \ref{Fig:2fsPulsesp} reveal that Rabi flopping produces a clear signature in the spectrum of resonance fluorescence of Gaussian pulses, which are becoming available by self-seeding at LCLS \cite{arXiv:1003.2548v1, *arXiv:1008.3036v1, Amann}. However, since shot-to-shot variations in pulse intensity and duration are anticipated, we investigate how the spectrum of resonance fluorescence is influenced by the presence of these fluctuations. For this purpose, we compute the energy spectrum of resonance fluorescence for a wide set of Gaussian pulses [Eq.~(\ref{eq:Gaussian_function})], by independently randomizing their duration and energy. The mean duration is chosen to be $\bar{\tau}_{\mathrm{G}} = 7\,\unit{fs}$ and the mean peak intensity is $\bar{I}_{\mathrm{G}} = 7\times10^{17}\,\mathrm{W/cm^2}$, giving a mean peak Rabi frequency of $\approx\,6\,\unit{eV}$. We compute the energy spectrum of resonance fluorescence for 500 different realizations of the driving pulses. Thereby, the duration and the intensity are random variables whose probability distribution is Gaussian with a variance of $20\%$ of the mean value. The resulting average resonance fluorescence spectrum is shown in Fig.~\ref{Fig:regular_averaged_spectrum}. It reveals that Rabi flopping is discernible even if the energy and duration of the pulse vary appreciably from shot to shot.

\subsection{SASE x-ray pulses}
\label{Chaotic x-rays}

\begin{figure}[!t]
\centering%
\includegraphics[width=0.875\linewidth, keepaspectratio]{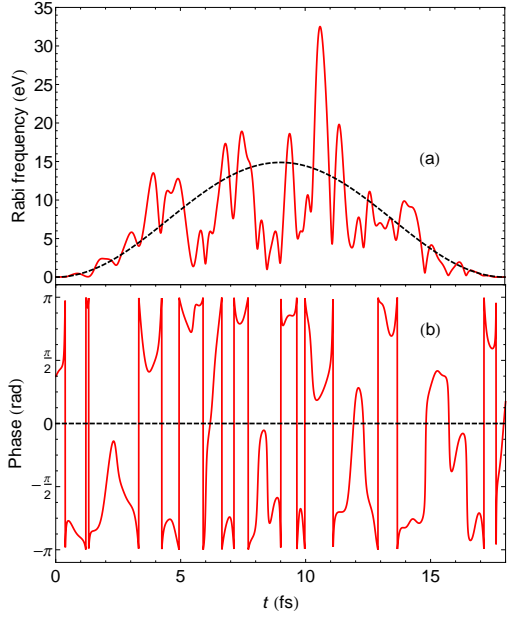}
\caption{(Color online)  (\textit{a}) The Rabi frequency $\varOmega_{\mathrm{R}}(t)$ induced by the amplitude of a SASE pulse and (\textit{b}) the phase $\varphi_{\mathrm{X}}(t)$ of the SASE pulse (red, solid lines) and their mean value (black, dashed lines). The mean pulse has a duration $\tau_{\mathrm{env}}= 6.5\,\unit{fs}$ and a peak intensity $I = 3.8\times 10^{18}\,\mathrm{W/cm^2}$. Its bandwidth is $\Delta\omega_{\mathrm{SASE}} = 6\,\unit{eV}$.}
\label{Fig:chaotic_pulse_time}
\end{figure}

\begin{figure}[!hbtp]
\centering%
\includegraphics[width=0.87\linewidth, keepaspectratio]{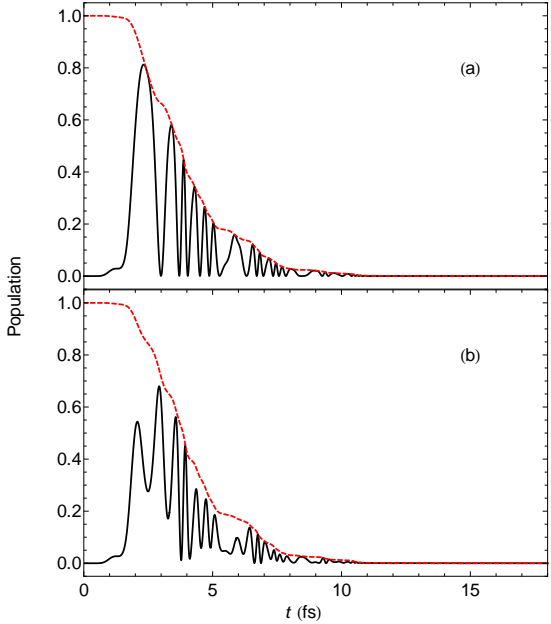}
\caption{(Color online)  Time evolution of a two-level system driven by a SASE pulse with the time-dependent Rabi frequency of Fig.~\ref{Fig:chaotic_pulse_time}\textit{a}. The phase is assumed to be (\textit{a}) constant, $\varphi_{\mathrm{X}}(t)=0$, or (\textit{b}) to be equal to the phase of Fig.~\ref{Fig:chaotic_pulse_time}\textit{b}. The red, dashed line shows the evolution of the total population $\rho_{11}(t) + \rho_{22}(t)$ [Eq.~(\ref{eq:density_matrix})]; the black, solid line represents the occupation of the excited state $\rho_{22}(t)$.}
\label{Fig:chaotic_population}
\end{figure}

\begin{figure}[!hbp]
\centering%
\includegraphics[width=0.87\linewidth, keepaspectratio]{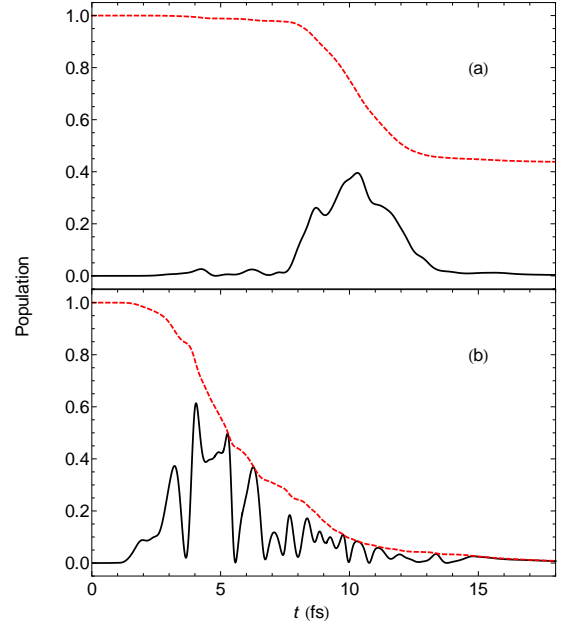}
\caption{(Color online)  Time evolution of a two-level system driven by SASE pulses generated with the PCM described in Appendix~\ref{Partial Coherent Method}. In both cases the pulses have a mean duration $\tau_{\mathrm{env}}= 6.5\,\unit{fs}$ and a bandwidth $\Delta\omega_{\mathrm{SASE}} = 6\,\unit{eV}$. The peak intensity is (\textit{a}) $I = 3.8\times 10^{15}\,\mathrm{W/cm^2}$ and (\textit{b})  $I = 8.8\times 10^{17}\,\mathrm{W/cm^2}$. The red, dashed line shows the evolution of the total population $\rho_{11}(t) + \rho_{22}(t)$ [Eq.~(\ref{eq:density_matrix})]; the black, solid line represents the occupation of the excited state $\rho_{22}(t)$.}
\label{Fig:chaotic_populationDifferentIntensities}
\end{figure}

After investigating resonance fluorescence with laserlike regular Gaussian pulses, we turn to the presently available SASE pulses at LCLS. The SASE light is modeled with the partial coherence method (PCM) \cite{Pfeifer:10, PhysRevA.82.041403}, whose details are discussed in Appendix~\ref{Partial Coherent Method}. The SASE pulses have a central photon energy which is tuned to the transition energy of $\text{Ne}^+$ of $848\,\unit{eV}$, with a bandwidth (FWHM of $|\tilde{\mathcal{E}}(\omega)|^2$) of $\Delta\omega_{\mathrm{SASE}} = 6\,\unit{eV}$. The envelope function $f(t)$ that we adopt [Eq.~(\ref{eq:envelopeyeah})] has FWHM duration $\tau_{\mathrm{env}} = 6.5\,\unit{fs}$. Further details are discussed in Appendix~\ref{Partial Coherent Method}.

In Fig.~\ref{Fig:chaotic_pulse_time}\textit{a} we display the time-dependent Rabi frequency [Eq.~(\ref{eq:omegaR})] $\varOmega_{\mathrm{R}}(t) = \wp\,\mathcal{E}_0(t)$ induced by the amplitude of a SASE pulse and in Fig.~\ref{Fig:chaotic_pulse_time}\textit{b} the phase $\varphi_{\mathrm{X}}(t)$ of a SASE LCLS pulse obtained with the PCM method. The mean Rabi frequency and phase are also given. In Fig.~\ref{Fig:chaotic_population}\textit{a} the time evolution of the population of the excited state and the total population of the two-level system are plotted if the phase of the pulse is supposed constant, $\varphi_{\mathrm{X}}(t) = 0$, and the spiky time-dependent Rabi frequency of Fig.~\ref{Fig:chaotic_pulse_time}\textit{a} is used to integrate the EOMs. In Fig.~\ref{Fig:chaotic_population}\textit{b} both the Rabi frequency and the phase of Fig.~\ref{Fig:chaotic_pulse_time} are used to integrate the EOMs. If the phase fluctuations are neglected, the decay of the system is slower; in both cases, due to the chaotic SASE pulse shape, the time evolution is very irregular. For the case displayed in Fig.~\ref{Fig:chaotic_population}\textit{a}, though, one can see the presence of complete oscillations in $\rho_{22}(t)$, reaching its minimum at $\rho_{22}(t) = 0$ and its maximum when $\rho_{11}(t) = 0$: this feature disappears when the phase fluctuations of Fig.~\ref{Fig:chaotic_pulse_time}\textit{b} are taken into account. 

In contrast to the case of a Gaussian pulse, one cannot extract from the time evolution of the system any clear relation to the pulse area. Nevertheless, one observes a relation between the Rabi frequency of Fig.~\ref{Fig:chaotic_pulse_time}\textit{a} and the frequency with which the population of the excited state $\rho_{22}(t)$ oscillates in Fig.~\ref{Fig:chaotic_population}. These oscillations, in fact, take place in a time interval which is shorter than the time characterizing the random fluctuations of Fig.~\ref{Fig:chaotic_pulse_time}. They are Rabi oscillations induced by the interaction with the intense driving field; as we show in Fig.~\ref{Fig:chaotic_populationDifferentIntensities}, if a pulse of similar bandwidth but of far lower intensity is used to excite the system, the time evolution of the atomic system displays slower oscillations, whose mean frequency increases at increasing intensities. We further notice that, because of photoionization, the increase in the intensity of the driving field reduces the actual decay time of the system: this emerges by comparing the graphs displayed in Fig.~\ref{Fig:chaotic_populationDifferentIntensities} with that of Fig.~\ref{Fig:chaotic_population}\textit{b}.

\begin{figure}[bt]
\centering%
\includegraphics[width=0.95\linewidth, keepaspectratio]{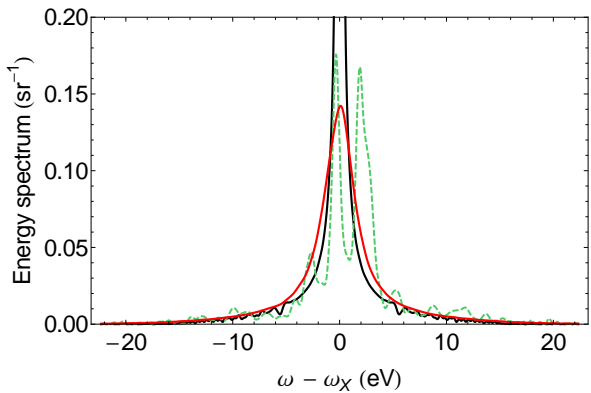}
\caption{(Color online)  Resonance fluorescence spectrum for SASE pulses. The black, dotted line shows the spectrum from a Gaussian pulse with FWHM duration $\tau_{\mathrm{G}} = 6.5\,\unit{fs}$ and peak intensity $I_{\mathrm{G}}= 3.8\times10^{18}\,\mathrm{W/cm^2}$. The red, solid line is the arithmetic mean over 1000 SASE pulses with average peak intensity $I_{\mathrm{G}}$, FWHM duration $\tau_{\mathrm{G}}$, and a bandwidth of $\Delta\omega_{\mathrm{SASE}} = 6\,\unit{eV}$. The green, dashed line is for the pulse in Fig.~\ref{Fig:chaotic_pulse_time}.}
\label{Fig:chaotic_spectrum}
\end{figure}

In Fig.~\ref{Fig:chaotic_spectrum} we display the resonance fluorescence spectrum from SASE x rays. To observe Rabi flopping we need the Rabi oscillations to occur within the coherence time of the pulse; for this reason, the intensity of the external electric field is chosen such that the maximum Rabi frequency is larger than the bandwidth $\Delta\omega_{\mathrm{SASE}}$ of the pulse itself. We look, in particular, at the emitted spectrum by averaging over 1000 independent SASE pulses. The tails appearing in the spectrum of Fig.~\ref{Fig:chaotic_spectrum} are nonvanishing contributions at frequencies higher than the bandwidth of the pulse itself. These tails would not appear if the field had equal bandwidth but lower intensity: they represent, therefore, a signature of the Rabi oscillations described in Fig.~\ref{Fig:chaotic_population}. These tails are also in good agreement with the spectrum emitted when a Gaussian transform-limited pulse of identical intensity and time duration---but clearly with much lower bandwidth---is used to excite the system. If the phase of the SASE pulse remained constant and only its amplitude displayed chaotic fluctuations, then the spectrum emitted after one single pulse would be symmetric; furthermore, the average spectrum would present a lower width, due to the absence of phase fluctuations. A clear observation of the tails of Fig.~\ref{Fig:chaotic_spectrum} and of the enlargement of the resonance fluorescence spectrum at increasing intensities might represent a possible way to detect Rabi flopping also at present SASE facilities.

Analogous conclusions had been drawn for the resonant Auger electron spectrum \cite{PhysRevA.77.053404}: the width of the resonant Auger electron line profile was expected to help in estimating the presence of Rabi oscillations in the system. Nonetheless, the very short coherence times at present XFEL facilities limited the actual experimental observability of this effect at LCLS \cite{PhysRevLett.107.233001}.

\begin{figure}[bt]
\centering%
\includegraphics[width=0.95\linewidth, keepaspectratio]{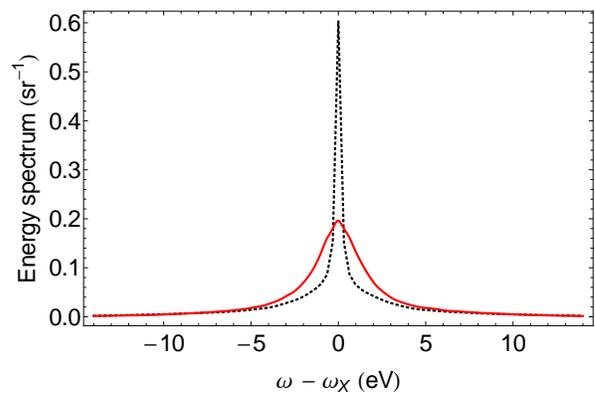}
\caption{(Color online)  Average resonance fluorescence spectrum over SASE pulses. The pulses have an average peak intensity of $I = 1.6\times10^{18}\,\mathrm{W/cm^2}$ and bandwidth $\Delta\omega_{\mathrm{SASE}} = 6\,\unit{eV}$. The red, solid line shows the average over SASE pulses with average FWHM duration of $\tau_{\mathrm{env}}= 6.5\,\unit{fs}$; the black, dotted line is associated with pulses of average FWHM duration of $\tau_{\mathrm{env}}= 2\,\unit{fs}$.}
\label{Fig:chaotic_comparison}
\end{figure}

As a last point, we study the dependence of the resonance fluorescence spectrum on the duration of the SASE pulse. In Fig.~\ref{Fig:chaotic_comparison} we plot the average spectrum emitted by $\text{Ne}^+$ cations when excited by an ultrashort pulse with peak intensity $I = 1.6\times10^{18}\,\mathrm{W/cm^2}$ and bandwidth $\Delta\omega_{\mathrm{SASE}} = 6\,\unit{eV}$. The results are obtained by averaging over spectra resulting from SASE pulses, respectively, with a FWHM duration of $\tau_{\mathrm{env}}= 6.5\,\unit{fs}$ and of $\tau_{\mathrm{env}}= 2\,\unit{fs}$. It is worth noticing a remarkable difference between different pulse durations. Naively, after the previous considerations, we would assume that the resonance fluorescence peak has a FWHM associated with the large bandwidth of the pulse $\Delta\omega_{\mathrm{SASE}} = 6\,\unit{eV}$. For the shortest pulses, though, the resonance fluorescence spectrum exhibits a higher central peak whose width is clearly lower than $\Delta\omega_{\mathrm{SASE}}$. The explanation is based on the same arguments that we used to describe the spectra depicted in Fig.~\ref{Fig:2fsPulsesp}\textit{a}, in which the post-x-ray-exposure decay results in a high Lorentzian peak of width given by the Auger decay width of the system. Analogously, for the ultrashort SASE pulses with $\tau_{\mathrm{env}} = 2\,\unit{fs}$ used in Fig.~\ref{Fig:chaotic_comparison}, the interaction with the pulse is shorter than the time needed by the system to completely decay; hence, at the end of the pulse, the probability of destruction of the system is about 90\%. The Auger decay which follows the interaction with an ultrashort SASE pulse implies, therefore, the high central peak in the resonance fluorescence spectrum shown in Fig.~\ref{Fig:chaotic_comparison}; for the same reason, its width is lower than the bandwidth of the pulse itself. A similar reduction of the FWHM was also observed in \cite{PhysRevLett.107.233001} in the Auger electron spectrum. In that case, by using the same \mbox{x-ray} pulse to create Ne$^+$ ions and to drive the $1s\,2p^{-1}\rightarrow 1s^{-1}\, 2p$ transition, the system could not completely Auger decay before the end of the pulse. The decay of the excited state with the natural decay time of the system turned out to dominate the observation.

\section{Conclusion}
\label{Conclusion}

In this paper we study theoretically the resonance fluorescence for intense ultrashort x rays. The fundamental role played by resonance fluorescence for the study of the quantum properties of light renders it a cornerstone of x-ray quantum optics. It also gives an alternative and more easily available point of view on resonant Auger decay in ultraintense x rays, whose study, at present, is not yet fully conclusive \cite{PhysRevLett.107.233001}. Therefore, we investigate the nonlinear phenomena of Rabi flopping, i.e., repeated cycles of stimulated emission and absorption of photons induced by the interaction with the ultraintense pulses from XFELs, and its signature in the resonance fluorescence spectrum at \mbox{x-ray} frequencies.

We develop a two-level model of resonance fluorescence whose time evolution is described by master equations which include the coherent interaction of the system with the classical \mbox{x-ray} field. All processes that destroy the system, namely, Auger decay and photoionization, are fully taken into account. We use our model to describe Ne$^+$ cations driven by an intense linearly polarized \mbox{x-ray} field tuned to the $1s\,2p^{-1}\rightarrow 1s^{-1}\, 2p$ transition at $848\,\unit{eV}$; the transition is well isolated, i.e., separated by more than 70 natural linewidths from the lowest lying Rydberg excitation, $1s\rightarrow 3p$ \cite{PhysRevLett.107.233001}. The intensity available at present \mbox{x-ray} FELs such as LCLS is sufficiently high to induce Rabi flopping at frequencies that compete with the rate of destruction of the system. The two-level approximation allows us to investigate the resonance fluorescence of photons associated with the transition to the state with $M_L = 0$ for two different scenarios. First, we consider SASE radiation from present XFELs; second, we explore resonance fluorescence from coherent Gaussian pulses which are becoming available via the use of self-seeding techniques at fourth-generation \mbox{x-ray} sources. The measurement of the spectra predicted in this paper need to take advantage of the polarization properties of the emitted light.

In the case of laserlike Gaussian pulses a clear signature of Rabi flopping is predicted. We further show that the observation of Rabi flopping persists even when intensity and duration of the pulse vary appreciably from shot to shot. For SASE pulses, even though Rabi flopping does not manifest itself as clearly as in the previous case, we predict the appearance of tails in the spectrum that might represent a good signature of Rabi oscillations in the atomic system. These tails would not appear if the system was excited by a less intense pulse of equally large bandwidth. In the case of the resonant Auger spectrum, however, the presently large bandwidth at LCLS represented a limit for the observation of analogous effects in the resonant Auger electron line \cite{PhysRevLett.107.233001} and the signature of Rabi flopping did not appear indistinguishably. Also in the case of resonance fluorescence the identification of Rabi flopping in the spectrum might be challenging. The amplitude of the aforementioned tails in which one is interested is predicted to be neither very high nor easily distinguishable. A much clearer signature is, however, identified for ions driven by Gaussian pulses, making the prospects with self-seeded LCLS very promising.

The results which have been presented motivate further experimental investigation of resonance fluorescence at XFEL facilities. In particular, the method which has been discussed here is a good candidate for further studies at hard x-ray frequencies. In the case of argon cations, for instance, the spectrum of resonance fluorescence, because of the higher fluorescence yield compared with that of neon, is more intense than the one predicted in this paper. By including the radiative decay width in the EOMs (\ref{eq:diffR}), the model can be used to study the resonance fluorescence spectrum of cations, e.g., argon, with higher fluorescence yield than the cations considered here. The basic features are discussed in this work and no qualitative differences are expected in heavier cations. Studies of resonance fluorescence in different atomic systems might require the use of a generalized formalism. The two-level approximation adopted in this paper, in fact, is not always sufficient to properly describe the atomic transitions of interest: in these cases multilevel systems have to be considered. In addition, in order to take into account the atomic properties for different \mbox{x-ray} transitions, a considerable amount of new atomic data is necessary, such as, for example, Auger pathways and decay rates. A detailed theoretical study of the atomic properties of the system would have to be implemented, motivating further research in this field.

Studies of resonance excitations followed by $K$-shell photoionization are receiving a lot of interest also for their potential applications in the biomedical sector \cite{Nahar20081951, doi:10.1021/jp904977z, doi:10.1021/jp905323y, theranostics}: Even though present facilities are not available yet for medical applications, studies of resonance fluorescence of $K$-shell transitions might also significantly contribute to the development of such applications of XFELs.

In addition, resonance fluorescence plays a crucial role in the study of the nonclassical properties of light, such as photon antibunching \cite{0022-3700-9-8-007, PhysRevA.13.2123, PhysRevLett.39.691, PhysRevA.18.2217} and squeezing \cite{PhysRevLett.47.709, PhysRevLett.49.136, WallsNature, drummond2004quantum}. Our study opens the \mbox{x-ray} regime up for quantum optical effects which can be investigated by means of ultraintense pulses now available at XFELs.

Finally, for nonstationary light, e.g., when the electric field has a pulse-shaped envelope, the study of the time-dependent power spectrum \cite{0022-3700-13-2-011, Eberly:77, PhysRevA.26.892, PhysRevA.37.1576, PhysRevLett.59.2149}, i.e., the time-dependent rate of detected photons, would allow one to investigate how the spectral properties of the fluorescent light evolve during the pulse. Even though such a power spectrum cannot be measured at present because of the ultrashort nature of XFEL pulses, with duration of the order of $10$-$100\,\unit{fs}$, and because of the lack of sufficiently fast detectors, the study of the time-dependent power spectrum might provide better understanding and further knowledge of the interaction between matter and x~rays.

\acknowledgments
 
The work of Z.H.~was supported by the Alliance Program of the Helmholtz Association (HA216/EMMI). C.B., E.P.K., S.H.S., and L.Y.~were supported by the Chemical Sciences, Geosciences, and Biosciences Division, Office of Basic Energy Sciences, Office of Science, U.S.~Department of Energy, under Contract No.~DE-AC02-06CH11357.


\appendix

\section{Polarization effects and measurement geometry}
\label{Polarization effects and measurement geometry}

We consider here the resonance fluorescence spectrum emitted by the two-level system displayed in Fig.~\ref{Fig:themodel} which is measured by rotating the detector around the \mbox{$y$ axis} of Fig.~\ref{Fig:experimental_setup}, i.e. the spectrum at point $\boldsymbol{r} = r\,\hat{\boldsymbol{e}}_r(\theta)$, with $\hat{\boldsymbol{e}}_r(\theta) = \cos\theta\,\hat{\boldsymbol{e}}_x + \sin\theta\,\hat{\boldsymbol{e}}_z$, where $\theta$ is the angle between $\hat{\boldsymbol{e}}_r(\theta)$ and the \mbox{$x$ axis}, lying in the $x-z$ plane. We further introduce the vector $\hat{\boldsymbol{e}}_{\theta}(\theta) =  - \sin\theta\,\hat{\boldsymbol{e}}_x + \cos\theta\,\hat{\boldsymbol{e}}_z$, which also lies in the $x-z$ plane and is orthogonal to $\hat{\boldsymbol{e}}_r(\theta)$; in this way, from (\ref{eq:polarization_in_model}) and (\ref{eq:emitted_field}), one has that 
$\hat{\boldsymbol{E}}^{+}(r\,\hat{\boldsymbol{e}}_r(\theta), \,t)= \hat{{E}}^{+}_{\theta}(r\,\hat{\boldsymbol{e}}_r(\theta), \,t)\,\hat{\boldsymbol{e}}_{\theta}(\theta) + \hat{{E}}^{+}_y(r\,\hat{\boldsymbol{e}}_r(\theta), \,t)\,\hat{\boldsymbol{e}}_y$,
with
\begin{equation}
\begin{aligned}
\hat{{E}}^{+}_{\theta}(r\,\hat{\boldsymbol{e}}_r(\theta), \,t) =\, &\frac{\wp\,\omega_{21}^2}{c^2 r }\,\Bigl[\cos\theta\,\hat{\sigma}_{1_02}(t') \Bigr.\\ 
& \Bigl.+ \sin\theta\,\frac{\hat{\sigma}_{1_+2}(t') - \hat{\sigma}_{1_-2}(t')}{\sqrt{2}}\Bigr]
\end{aligned}
\end{equation}
and
\begin{equation}
\hat{{E}}^{+}_y(r\,\hat{\boldsymbol{e}}_r(\theta), \,t) =\, \frac{\uimm}{\sqrt{2}}\,\frac{\wp\,\omega_{21}^2}{c^2 r }\, \left(\hat{\sigma}_{1_+2}(t')+\hat{\sigma}_{1_-2}(t')\right),
\end{equation}
with $t' = t-r/c$. The autocorrelation function (\ref{eq:autocorrelation}) is $G^{(1)}(t_1,t_2, r\,\hat{\boldsymbol{e}}_r(\theta)) = G^{(1)}_{\theta}(t_1,t_2, r\,\hat{\boldsymbol{e}}_r(\theta)) + G^{(1)}_y(t_1,t_2, r\,\hat{\boldsymbol{e}}_r(\theta))$, with
\begin{equation}
\begin{aligned}
& G^{(1)}_{\theta}(t_1,t_2, r\,\hat{\boldsymbol{e}}_r(\theta)) =\,  \mathcal{I}(r)\,\Bigl[\cos^2\theta\,\langle \hat{\sigma}_{21_0}(t_1')\,\hat{\sigma}_{1_02}(t_2')\rangle \Bigr.\\
& + \Bigl.\frac{\sin^2\theta}{2}\,\bigl(\langle \hat{\sigma}_{21_-}(t_1')\,\hat{\sigma}_{1_-2}(t_2')\rangle + \langle \hat{\sigma}_{21_+}(t_1')\,\hat{\sigma}_{1_+2}(t_2')\rangle\bigr)\Bigr]
\end{aligned}
\end{equation}
and
\begin{equation}
\begin{aligned}
G^{(1)}_{y}(t_1,t_2, r\,\hat{\boldsymbol{e}}_r(\theta)) =\, \mathcal{I}(r)\,\frac{1}{2}\,& \bigl(\langle \hat{\sigma}_{21_-}(t_1')\,\hat{\sigma}_{1_-2}(t_2')\rangle \bigr.\\
&\bigl. + \langle \hat{\sigma}_{21_+}(t_1')\,\hat{\sigma}_{1_+2}(t_2')\rangle\bigr),
\end{aligned}
\end{equation}
where $\mathcal{I}(r)$ is defined in (\ref{eq:Ir}) and the application of the quantum regression theorem allows one to show that the crossterms $\langle \hat{\sigma}_{2i}(t_1')\,\hat{\sigma}_{j2}(t_2')\rangle$, with $i,\,j\in\{1_+,\,1_-, 1_0\}$, $i\neq j$, vanish for any $t_1'$ and $t_2'$. An analogous spatial dependence can be displayed also in the resonance fluorescence spectrum $S(\omega,\,r\,\hat{\boldsymbol{e}}_r(\theta)) = S_{\theta}(\omega,\,r\,\hat{\boldsymbol{e}}_r(\theta)) + S_y(\omega,\,r\,\hat{\boldsymbol{e}}_r(\theta))$. 

In conclusion, we notice that the photons emitted in transitions to the two undriven states $|1_{\pm}\rangle$ can both be polarized along the axes $\hat{\boldsymbol{e}}_y$ and $\hat{\boldsymbol{e}}_{\theta}(\theta)$. Conversely, the photons spontaneously emitted to the state $|1_0\rangle$ are exclusively polarized along the axis $\hat{\boldsymbol{e}}_{\theta}(\theta)$ and their intensity, varying in space as $\cos^2\theta$, is maximized for $\theta = 0, \,\pi$. For the same angles the intensity of the photons that are emitted in transitions to the two undriven states $|1_{\pm}\rangle$ and which are linearly polarized along $\hat{\boldsymbol{e}}_{\theta}(\theta)$ vanishes. This motivates our choice throughout the paper of studying the spectrum of resonance fluorescence for $\theta = 0$, $\hat{\boldsymbol{e}}_r = \hat{\boldsymbol{e}}_x$ and $\hat{\boldsymbol{e}}_{\theta} = \hat{\boldsymbol{e}}_z$. Polarization-dependent detection of the resonance fluorescence spectrum can take advantage of the properties just presented.

\section{Partial Coherence Method}
\label{Partial Coherent Method}

We use the partial coherence method (PCM) introduced in Ref.~\cite{Pfeifer:10} to generate random realizations of the temporal shape of SASE XFEL pulses, whose knowledge is an important prerequisite for meaningful investigations of nonlinear x ray-matter interaction \cite{PhysRevA.77.053404}. Those parameters which can be measured at present XFELs, such as the average spectral intensity and the pulse duration, are taken into account as input parameters \cite{PhysRevA.82.041403}. The PCM is used to generate non-transform-limited pulses, with a coherence time lower than the average FWHM duration of the pulse and with significant fluctuations in the pulse shape from shot to shot. 

The pulses are generated starting from their frequency representation $\tilde{\mathcal{E}}(\omega)$, whose amplitude is given by the average spectral intensity of the pulse. If the phase of $\tilde{\mathcal{E}}(\omega)$ was constant, then by Fourier transform one would obtain a transform-limited pulse. In order to generate a SASE pulse, we let the spectral phase vary in $[-\pi,\,\pi[$.

The PCM models the classical electric field $\mathcal{E}(t)$ [Eq.~(\ref{eq:classical_field})]. We introduce the complex electric field \cite{diels2006ultrashort} $E^{\pm}(t) = \frac{1}{2}\mathcal{E}_0(t)\,\eu^{\mp\uimm[\varphi_{\mathrm{X}}(t) + \omega_{\mathrm{X}}t]}$ and the complex field envelope $\tilde{\mathcal{E}}(t) = \frac{1}{2}\mathcal{E}_0(t)\,\eu^{-\uimm\varphi_{\mathrm{X}}(t)}$, such that $\mathcal{E}(t) = \tilde{\mathcal{E}}(t)\,\eu^{-\uimm\omega_{\mathrm{X}}t } + \tilde{\mathcal{E}}^*(t)\,\eu^{\uimm\omega_{\mathrm{X}}t }$. It follows that \cite{diels2006ultrashort}
\begin{equation}
|{\mathcal{E}}(t)|^2 = 2 |E^{\pm}(t) |^2 = 2 |\tilde{\mathcal{E}}(t)|^2 = \frac{|\mathcal{E}_0(t)|^2}{2}.
\label{eq:intensities_defs}
\end{equation}

We define the Fourier transform of $\tilde{\mathcal{E}}(t)$ as
\begin{equation}
\tilde{\mathcal{E}}(\omega) = \int_{-\infty}^{+\infty} \tilde{\mathcal{E}}(t)\eu^{\uimm\omega t}\,\diff t = |\tilde{\mathcal{E}}(\omega)|\,\eu^{-\uimm\phi(\omega)}
\label{eq:Eomega}
\end{equation}
and from Parseval's theorem it follows that
\begin{equation}
\int_{-\infty}^{+\infty}|\tilde{\mathcal{E}}(t)|^2\,\diff t = \frac{1}{2\pi} \int_{-\infty}^{+\infty}|\tilde{\mathcal{E}}(\omega)|^2\,\diff \omega.
\label{eq:Parseval}
\end{equation}
Analogously one can define ${\mathcal{E}}(\omega)$ and $E^+(\omega)$ as the Fourier transforms of ${\mathcal{E}}(t)$ and $E^+(t)$ respectively. One notices that $E^+(\omega) = \tilde{\mathcal{E}}\left(\omega-\omega_{\mathrm{X}}\right)$ and therefore $\mathcal{E}(\omega) = \tilde{\mathcal{E}}\left(\omega-\omega_{\mathrm{X}}\right) + \tilde{\mathcal{E}}\left(\omega+\omega_{\mathrm{X}}\right)$, so that, from Parseval's theorem (\ref{eq:Parseval}), one finds in agreement with (\ref{eq:intensities_defs}) that
\begin{equation}
\int_{-\infty}^{+\infty}|{\mathcal{E}}(t)|^2\,\diff t  = \frac{1}{2\pi} \int_{-\infty}^{+\infty}2\,|\tilde{\mathcal{E}}(\omega)|^2\,\diff \omega.
\end{equation}

The average spectral intensity of a SASE pulse is modeled here---close to measured spectral intensities---as a Gaussian function, so that
\begin{equation}
|\tilde{\mathcal{E}}(\omega)|^2 = |\tilde{\mathcal{E}}_{\mathrm{sp,max}}|^2 \eu^{-(\omega^2/\varOmega^2)},
\label{eq:spectral_energy}
\end{equation}
whose FWHM is $\Delta\omega_{\mathrm{SASE}} = 2\varOmega\sqrt{\ln2}$. The FWHM duration of the squared modulus of the inverse Fourier transform of $|\tilde{\mathcal{E}}(\omega)|$ \cite{diels2006ultrashort}, which is here also a Gaussian function, is $\tau_{\mathrm{SASE}} = 4\ln2/\Delta\omega_{\mathrm{SASE}}$. It follows that
\begin{equation}
|{\mathcal{E}}(\omega)|^2 = |\tilde{\mathcal{E}}_{\mathrm{sp,max}}|^2 \Bigl(\eu^{-[(\omega-\omega_{\mathrm{X}})^2/\varOmega^2]} + \eu^{-[(\omega+\omega_{\mathrm{X}})^2/\varOmega^2]}\Bigr).
\end{equation}

The average spectral intensity, though, does not provide any information about the spectral phase of the pulse. In analogy to the phase retrieval in x-ray crystallography \cite{Als-Nielsen:EM-01}, the knowledge of the spectral amplitude is not sufficient to completely determine the temporal shape of the pulse via inverse Fourier transform. In the PCM approximate phase retrieval is achieved by assuming initially a random frequency-dependent spectral phase varying in $[-\pi,\,\pi[$. 

We define a discrete spectral component of the electric field $\tilde{\mathcal{E}}(\omega_i) = |\tilde{\mathcal{E}}(\omega_i)|\,\eu^{-\uimm\phi_i}$, with a sampling interval $|\omega_{i+1} - \omega_{i}| \ll \Delta\omega_{\mathrm{SASE}}$. $\phi_i$ are random numbers in $[-\pi,\,\pi[$. The discrete inverse Fourier transform of $\tilde{\mathcal{E}}(\omega_i)$ provides the time-dependent discrete field $\mathcal{R}(t_j)$. 

The complex function $\mathcal{R}(t)$, obtained by interpolating $\mathcal{R}(t_j)$, spans an infinitely long interval in time because of the fluctuating $\phi_i$. To generate SASE pulses of finite duration, we multiply $\mathcal{R}(t)$ by a temporal filter function $f(t)$. This function is non-zero only in a finite domain and the FWHM duration of $|f(t)|^2$ is $\tau_{\mathrm{env}}$. The finite duration of FEL pulses is determined by the electron bunch duration \cite{1367-2630-12-3-035021} and is usually measured. All together, we approximate the complex electric field by
\begin{equation}
\tilde{\mathcal{E}}(t) = \frac{1}{2}\mathcal{E}_0(t)\,\eu^{-\uimm\varphi_{\mathrm{X}}(t)} = \mathcal{R}(t)f(t).
\end{equation}

Along the way, we notice that the inverse Fourier transform of $\mathcal{R}(t)f(t)$ is given by the convolution of the respective inverse Fourier transforms $\tilde{\mathcal{E}}(\omega)$ and $\tilde{f}(\omega)$. $\tilde{\mathcal{E}}(\omega)$ has a random fluctuating phase $\phi(\omega)$, whereas $|\tilde{f}(\omega)|^2$ has a spectral FWHM $\Delta\omega_{\mathrm{env}}$ related to the inverse of $\tau_{\mathrm{env}}$. Hence, the spectral amplitude of a single pulse generated with the PCM also displays a spiky structure, where the average FWHM frequency of each spike is about $\Delta\omega_{\mathrm{env}}$ \cite{1367-2630-12-3-035021}. In addition, since the average value of $\phi(\omega)$ is 0, the average spectral amplitude results from the convolution of $|\tilde{\mathcal{E}}(\omega)|$ and $\tilde{f}(\omega)$ and, because $\tau_{\mathrm{env}}\gg \tau_{\mathrm{SASE}}$, the width of $\tilde{f}(\omega)$ is much narrower than the width of $|\tilde{\mathcal{E}}(\omega)|$. Consequently, the convolution 
$$\int_{-\infty}^{+\infty}{|\tilde{\mathcal{E}}(\omega-\omega')|\tilde{f}(\omega')\,\diff\omega'}\approx |\tilde{\mathcal{E}}(\omega)|, $$
i.e., the multiplication by the envelope function $f(t)$ does not significantly affect the average spectral intensity of $\tilde{\mathcal{E}}(t)$.

To generate SASE pulses for this paper (Fig.~\ref{Fig:chaotic_pulse_time}) we use the envelope function 
\begin{equation}
f(t) = \left\{ 
\begin{aligned}
&f_0 \cos^2( {\pi t}/{T}) &\ \ \ \ &\text{if $|t|\leq T/2$}\\
&0                                       &\ \ \ \ &\text{if $|t|> T/2$}
\end{aligned}
\right.
\label{eq:envelopeyeah}
\end{equation}
with $T ={\pi \tau_{\mathrm{env}}}/{(2\arccos{\sqrt[4]{1/2}})}$ and $\tau_{\mathrm{env}}=6.5\,\unit{fs}$, defined as the FWHM duration of $|f(t)|^2$ \cite{0953-4075-42-23-235101}. The Fourier transform of $f(t)$ is
\begin{equation}
\tilde{f}(\omega) =\frac{T}{2}\ f_0\frac{\sinc\left(\frac{\omega T}{2}\right)}{1-\left(\frac{\omega T}{2\pi}\right)^2}.
\end{equation}
Then, $\Delta\omega_{\mathrm{env}} \approx 2.41/\tau_{\mathrm{env}}$ is the FWHM of $|\tilde{f}(\omega)|^2$. One notices that, for $\tau_{\mathrm{env}}= 6.5\,\unit{fs}$ and $\Delta\omega_{\mathrm{SASE}}=6\,\unit{eV}$, one has $\Delta\omega_{\mathrm{env}} = 0.24\,\unit{eV}\ll \Delta\omega_{\mathrm{SASE}} = 6\,\unit{eV}$.

Alternative approaches have also been developed and adopted, e.g., in \cite{Vannucci:80,  PhysRevA.76.033416}. In these cases, the electric field is written as a Fourier series in time domain
\begin{equation}
 \mathcal{E}(t) = \sum_{k=-\infty}^{\infty} a_k\,\cos(\omega_k t) + b_k\,\sin(\omega_k t),
\end{equation}
where the real coefficients $a_k$ and $b_k$ are independent zero-mean Gaussian random variables. Basically, this represents only a different description of $\tilde{\mathcal{E}}(\omega)$ compared with the PCM.

%

\end{document}